*galaxies* MDPI

*Review*

# Very High-Energy Emission from the Direct Vicinity of Rapidly Rotating Black Holes

**Kouichi Hirotani**

Academia Sinica, Institute of Astronomy and Astrophysics,
AS/NTU Astronomy-Mathematics Building, No.1, Sec. 4, Roosevelt Rd, Taipei 10617, Taiwan;
hirotani@asiaa.sinica.edu.tw; Tel.: +886-2-3365-5406

**Abstract:** When a black hole accretes plasmas at very low accretion rate, an advection-dominated accretion flow (ADAF) is formed. In an ADAF, relativistic electrons emit soft gamma-rays via Bremsstrahlung. Some MeV photons collide with each other to materialize as electron-positron pairs in the magnetosphere. Such pairs efficiently screen the electric field along the magnetic field lines, when the accretion rate is typically greater than 0.03–0.3% of the Eddington rate. However, when the accretion rate becomes smaller than this value, the number density of the created pairs becomes less than the rotationally induced Goldreich–Julian density. In such a charge-starved magnetosphere, an electric field arises along the magnetic field lines to accelerate charged leptons into ultra-relativistic energies, leading to an efficient TeV emission via an inverse-Compton (IC) process, spending a portion of the extracted hole's rotational energy. In this review, we summarize the stationary lepton accelerator models in black hole magnetospheres. We apply the model to super-massive black holes and demonstrate that nearby low-luminosity active galactic nuclei are capable of emitting detectable gamma-rays between 0.1 and 30 TeV with the Cherenkov Telescope Array.

## 1. Introduction

It is commonly accepted that every active galaxy harbors a supermassive black hole (BH), whose mass typically ranges between $10^6$ $M_\odot$ and $10^{9.5}$ $M_\odot$ in its center (e.g., [1–4]; see also [5,6] for reviews). Compelling evidence of a supermassive BH was found by observing the line emission from water masers around the central region of galaxy NGC 4258 [7]. Moreover, evidence of supermassive BHs at individual galactic centers are shown by the tight correlation between the black hole mass and the velocity dispersion or the bulge mass (e.g., [8–10]), which has been confirmed for the measurements with reverberation mapping (e.g., [11]).

A likely mechanism for powering such an active galactic nucleus (AGN) is the release of the gravitational energy of accreting plasmas [12] or the electromagnetic extraction of the rotational energy of a rotating supermassive BH [13]. The latter mechanism, which is called the Blandford– Znajek (BZ) mechanism, works only when there is a plasma accretion, because a BH cannot have its own magnetic moment (e.g., [14]). As long as the magnetic field energy is in a rough equipartition with the gravitational binding energy of the accreting plasmas, both mechanisms contribute comparably in terms of luminosity. The former mechanism is supposed to power the mildly relativistic winds that are launched from the accretion disks [15–17]. There is, however, growing evidence that relativistic jets are energized by the latter BZ mechanism through numerical simulations [18–20] (see also [21] for an ergospheric disk jet model). Indeed, general relativistic (GR) magnetohydrodynamic (MHD) models show the existence of collimated and magnetically dominated jets in the polar regions [22–24], whose structures are similar to those in the force-free models [25-27]. Since the centrifugal-force barrier prevents plasma accretion toward the rotation axis, the magnetic energy density dominates the plasmas' rest-mass energy density in these polar funnels.

Within such a nearly vacuum, polar funnel, electron–positron pairs are supplied via the collisions of MeV photons emitted from the equatorial, accreting region. For example, when the mass accretion rate is typically less than 1% of the Eddington rate, the accreting plasmas form an advection-dominated accretion flow (ADAF), emitting radio to infrared photons via the synchrotron process and MeV photons via free–free and inverse-Compton (IC) processes [28,29]. Particularly, when the accretion rate becomes much less than the Eddington rate, the ADAF MeV photons can no longer sustain a force-free magnetosphere, which inevitably leads to the appearance of an electric field, $E_\parallel$, along the magnetic field lines in the polar funnel. In such a vacuum gap, we can expect that the BZ power may be partially dissipated as particle acceleration and emission near the central engine, in the same manner as in pulsar outer gap (OG) model. In what follows, we summarize this vacuum gap model in BH magnetospheres.

## 2. The Pulsar Outer Gap Model

The Large Area Telescope (LAT) aboard the *Fermi* space gamma-ray observatory has detected pulsed signals in high-energy (HE) (0.1 GeV–10 GeV) gamma-rays from more than 200 rotation-powered pulsars [30]. Among them, 20 pulsars exhibit pulsed signals above 10 GeV, including 10 pulsars up to 25 GeV and other 2 pulsars above 50 GeV. Moreover, more than 99% of the LAT-detected young and millisecond pulsars exhibit phase-averaged spectra that are consistent with a pure-exponential or a sub-exponential cut off above the cut-off energies at a few GeV. What is more, 30% of these young pulsars show sub-exponential cut off, a slower decay than the pure-exponential functional form. These facts preclude the possibility of emissions from the inner magnetosphere as in the polar-cap scenario [31–35], which predicts super-exponential cut off due to magnetic attenuation. That is, we can conclude that the pulsed gamma-ray emissions are mainly emitted from the outer magnetosphere that is close to the light cylinder, whose distance from the rotation axis is given by the so-called special-relativistic "light cylinder radius", $c/\Omega_F$, where $c$ denotes the speed of light and $\Omega_F$ the angular frequency of magnetic field rotation.

In a pulsar magnetosphere, the rotational energy of the neutron star is extracted by the magnetic torque and transferred outward as a Poynting flux. Most of such extracted energy is dissipated at large distances such as in the pulsar wind nebula. However, a small portion (typically 0.1–20%) can be dissipated as particle acceleration and resultant radiation in the outer magnetosphere, showing pulsed, incoherent emissions. Thus, the luminosity of a magnetospheric particle accelerator (i.e., a gap) does not exceed the neutron star's spin-down luminosity.

One of the main scenarios of such outer-magnetospheric emissions is the OG model [36–47]. In an OG, an electric field, $E_\parallel$, is exerted along the local magnetic field line near the null-charge surface, on which the rotationally induced Goldreich–Julian (GJ) charge density vanishes due to the convex geometry of the dipolar-like magnetic field of the neutron star. The $E_\parallel$ accelerate electrons and positrons into ultra-relativistic energies, leading to the emission of HE photons typically between 10 MeV and 10 GeV via the synchro-curvature process [48,49]. Electrons and positrons (which are referred to as leptons in this review) are accelerated by the magnetic-field aligned electric field, $E_\parallel$, and saturate at Lorentz factors typically below $10^{7.5}$ due to the radiation drag of the synchro-curvature process. Because of a superposition of the synchro-curvature emission from different places in the gap, a power-law spectrum can be obtained below the cut off, which appears typically at a few GeV [50]. In addition, if the system becomes non-stationary, the linear acceleration may play an important role, provided that the acceleration takes place within a short length scale, $R_c/\gamma$, where $R_c$ denotes the curvature radius of the 3-D particle motion, and $\gamma$ the lepton Lorentz factor. In this case, the resulting spectrum will show a power-law energy dependence blow the cutoff [51], in the same manner as in the jitter radiation in the synchrotron process.

This successful pulsar OG scenario was applied to BH magnetospheres, which will be described in the rest of this review.

## 3. The Black Hole Gap Model

*3.1. Stationary Gap Models*

Following the successful pulsar OG model, Beskin et al. applied it to BH magnetospheres for the first time [52]. It was shown that an efficient pair-production cascade can take place near the null-charge surface, where the GJ charge density vanishes due to the space–time frame-dragging around a rotating BH. Then it was demonstrated that the stationary gap solutions can be obtained from the set of Maxwell–Boltzmann equations in a consistent manner with the gap closure condition [53]. However, they considered mass accretion rates that are a good fraction of the Eddington accretion rate. Thus, the solved gap width, $w$, along the magnetic field lines, became much less than the gravitational radius, $r_g = GM/c^2$, where $M$ designates the BH mass, $c$ the speed of light, and $G$ the gravitational constant. In what follows, we adopt the geometrized unit, putting $c = G = 1$. Because of the small gap width, $w \ll M$, the exerted $E_\parallel$ is found to be very small compared to the magnetic-field strength, $B$, in a Gaussian unit; as a result, the gap luminosity, $L_{\text{gap}}$, became negligibly small compared to the Eddington luminosity. For a pure hydrogen gas, the Eddington luminosity becomes $L_{\text{Edd}} = 1.25 \times 10^{47} M_9$ erg s$^{-1}$, where $M_9 \equiv M/10^9 \, \text{M}_\odot$.

On these grounds, to achieve bright gap emissions, we began to consider spatially extended gaps, which can be possible if the soft photon field is weak enough so that the photon–photon pair-production mean-free path becomes comparable to or greater than the gravitational radius, $r_g = M$. For example, when the mass accretion rate is typically less than 1% of the Eddington rate, the accreting plasmas form an ADAF, emitting radio to infrared photons via synchrotron process and MeV photons via free–free and IC processes (§ 4). Under such a low accretion environment, gap-emitted TeV photons do not efficiently collide with the soft photons, leading to an un-screened, spatially extended gap, which has a much greater electric potential drop, and hence $L_{\text{gap}}$ than the denser soft-photon-field case.

In this context, Neronov and Aharonian [54] examined a BH gap emission and applied it to the central engine of M87, a nearby low luminosity AGN, which hosts a BH with mass $M \approx (3.2–6.6) \times 10^9 \, \text{M}_\odot$ [55–58]. They assumed that the gap is extended, $w \approx 2M$ (= Schwarzschild radius), that the lepton number density, $N_\pm$, is comparable to the typical GJ number density, $N_{\text{GJ}} \sim \Omega_F B/(2\pi c e)$, where $e$ denotes the magnitude of the charge on the electron, and that the magnetic field strength $B$ becomes comparable to the equipartition value,

$$B_{eq} \approx 4 \times 10^4 \, \dot{m}^{1/2} M_9^{-1/2}, \qquad (1)$$

which is obtained when the magnetic buoyancy balances the disk gravity; $\dot{m} \equiv \dot{M}/\dot{M}_{\text{Edd}}$ refers to the dimensionless accretion rate near the horizon; and $\dot{M}$ denotes the mass accretion rate. We have $L_{\text{Edd}} = \eta \dot{M}_{\text{Edd}} c^2$, where the radiation efficiency can be estimated as $\eta \approx 0.1$. They also assumed that the magnetic-field-aligned electric field is comparable to the perpendicular component,

$$E_\parallel \approx E_\perp \approx \frac{\varpi(\Omega_F - \omega)}{c} B \approx \frac{aB_H}{2M}, \qquad (2)$$

where this denotes the distance from the rotation axis, with the spacetime dragging angular frequency due to BH's rotation, $B_H$ does the magnetic field strength evaluated at the horizon. In the right-most near equality, we evaluate quantities near the horizon, where a gap is formed. Because of these four assumptions (i.e., $w \approx 2M$, $N_\pm \approx N_{\text{GJ}}$, $B \approx B_{eq}$, $E_\parallel \approx E_\perp$), $L_{\text{gap}}$ becomes comparable to the BZ power, $L_{\text{BZ}}$, which corresponds to the spin-down luminosity of pulsars. Under these assumptions, they demonstrated that the observed very-high-energy (VHE) emission of M87 can be explained as the IC emission of ultra-relativistic leptons accelerated in a BH gap. See section 7.1 for a brief summary of the VHE observations of M87 and other a few non-blazer AGNs.

Subsequently, Levinson and Rieger applied the vacuum gap model to M87 and found that the lepton density is strongly dependent on the accretion rate [59],

$$N_{\pm} = 3 \times 10^{11} \dot{m}^4 M_9^{-1} \text{ cm}^{-3}, \quad (3)$$

and that $L_{\text{gap}} \approx L_{\text{BZ}}$ holds if $w$ is a good fraction of $2M$. Considering the ADAF photon field near the BH, and assuming $w > 0.1 \times 2M$, $N_{\pm} \approx N_{\text{GJ}}$, $B \approx B_{\text{eq}}$ and $E_{\parallel} \approx (w/2M)^2 E_{\perp}$, they demonstrated that the VHE luminosity observed with Imaging Atmospheric Cherenkov Telescopes (IACTs) (7.1) can be reproduced by the BH gap model. They also argued that the gap activity may be intermittent, because the strong dependence of $N_{\pm}$ on $\dot{m}$ could lead to the disappearance of the gap due to moderate changes in the accretion rate.

Then Brodrick and Tchekhovskoy [60] considered a thick gap, $w \approx 2M$, and an equipartition magnetic field strength, $B \approx B_{\text{eq}}$, but adopted $N_{\pm} \approx 10^5 N_{\text{GJ}}$ and $E_{\parallel} \gg E_{\perp}$. They showed that the stagnation surface (§ 5.3) is a natural site of gap formation.

More recently, Hirotani and Pu [61] considered a one-dimensional BH gap, solving $w$, $N_{\pm}$, and $E_{\parallel}$ from the set of the inhomogeneous part of the Maxwell equations (i.e., the Poisson equation) for the non-corotational potential, lepton density and velocity at each position, and the radiative transfer equation (i.e., instead of assuming $w$, $N_{\pm}$, and $E_{\parallel}$, adopting $B \approx B_{\text{eq}}$. Then Hirotani et al. (2016) solved the same set of Maxwell–Boltzmann equations in the two-dimensional (2D) poloidal plane, assuming a mono-energetic approximation for the lepton distribution functions [62]. Subsequently, Hirotani et al. (2017) considered an inhomogeneous soft photon field to examine the γ-ray emission properties of supermassive BHs, solving the distribution functions of the accelerated leptons explicitly from their Boltzmann equations, discarding the mono-energetic approximation [63]. They showed that the accelerated leptons emit copious photons via IC processes between 0.1 and 30 TeV for a distant observer, and that these IC fluxes will be detectable with IACTs such as the Cherenkov Telescope Array (CTA), provided that a low-luminosity active galactic nucleus is located within 1 Mpc for a million-solar-mass central BH or within 30 Mpc for a billion-solar-mass central BH. Lin et al. then applied this method to stellar-mass BHs and compare the prediction with the high-energy (HE) observations, re-analyzing the archival Fermi/LAT data [64]. In addition, Song et al. demonstrated that the gap emission of an aligned rotator is enhanced along the rotation axis if the BH is nearly maximally rotating (i.e., $a \rightarrow M$) [65], because the magnetic fluxes concentrate polewards as $a \rightarrow M$ due to the magnetic pinch effect [24,66]. More recently, Hirotani et al. (2018a,b) investigated if stellar-mass BHs can emit detectable HE and VHE emissions when they encounter dense molecular clouds (6) [67,68]. In these stationary analysis [61–65,67,68], They found $L_{\text{gap}} \approx (10^{-1} - 10^{-4}) L_{\text{BZ}}$, depending on the accretion rate. The BZ power becomes [13,23]:

$$L_{BZ} = \Omega_F (\omega_H - \Omega_F) B_{\perp}^2 r_H^4, \quad (4)$$

where $\omega_H$ denotes the BH's rotational angular frequency, $r_H$ does the horizon radius, and $B_{\perp}$ does the strength of the magnetic field component perpendicular to the horizon. The BZ power maximizes when $\Omega_F = 0.5 \omega_H$. For a super-massive BH, the BZ power takes the typical value, $L_{BZ} \approx 10^{45} (a/M)^2 B_4^2 M_9^2$ erg/s, where $B_4 \equiv B/10^4$ G, and $a/M$ denotes the dimensionless BH spin, which vanishes for Schwarzschild BHs and becomes 1 for maximally rotating Kerr BHs. Exactly speaking, Equation (4) is obtained in the slow rotating limit, $a \ll M$ to the second order, $(a/M)^2$. For rapidly rotating BHs, the BZ power is obtained by Tanabe and Nagataki [69] to the fourth order, and by Tchekhovskoy et al. [24] to the sixth order. The higher-order corrections suppress the BZ power and a flattening occurs as a function of $(a/M)^2$ for extremely rotating BHs: for example, equation (4) over-predicts the BZ power when $a > 0.95M$ for geometrically thin disks.

*3.2. Force-Free Magnetosphere and the Necessity of a Gap*

For the BZ process to work, there should exist a global electric current in the magnetosphere. However, plasmas flow inwards inside the inner light surface and flow outwards outside the outer light surface (e.g., [70]). Thus, plasmas should be continuously replenished between the two light surfaces (see §5.2 for details) so that the BZ process may continuously work. If the created pair density, $N_\pm$, exceeds $N_{GJ}$, $E_\parallel$ will be quickly screened out by the charge redistribution and the magnetosphere is kept force-free. For $N_\pm > N_{GJ}$ to be realized, the soft photon density should be large enough near the event horizon. In the case of an ADAF, the density of the pairs created via the collisions of the ADAF-emitted MeV photons, becomes [59]:

$$N_\pm = 3\times 10^{11} \dot{m}^4 M_9^{-1} \text{ cm}^{-3}. \tag{5}$$

Comparing with $N_{GJ} \sim \Omega_F B/(2\pi ce)$, we find [59]:

$$\frac{N_\pm}{N_{GJ}} = 0.06 \left(\frac{\dot{m}}{10^{-4}}\right)^{7/2} M_9^{1/2} \tag{6}$$

Thus, if the accretion satisfies:

$$\dot{m} > 2.2\times 10^{-4} M_9^{-1/7} \tag{7}$$

the magnetosphere is kept force-free. In this case, there appears no gap. On the other hand, if the accretion rate stays below this value, the charge-starved magnetosphere inevitably has a gap, which may be either stationary or non-stationary.

Indeed, there is a growing consensus that the radio polarimetry in millimeter–submillimeter wavelengths provides useful diagnostics to infer the accretion rate of ADAF near the central engine. Through the Faraday rotation measure of the linear polarization, it is suggested that the dimensionless accretion rate becomes $\dot{m}$ =(1−10)×10$^{-6}$ within the radius $r < 200\,M$ (=*100* Schwarzschild radii) for Sgr A* [71–73] and $\dot{m} < 10^{-4}$ within $r < 42\,M$ for M87 [74]. In addition, the jet luminosity (~2 × 10$^{42}$ ergs s$^{-1}$) of radio galaxy IC 310 [75] shows that the time-averaged accretion rate is $\dot{m}$ ~10$^{-4}$ [76]. For low luminosity AGNs like M87 and IC 310 to sustain large-scale jets under such small accretion rates, sufficient pair creation should be taking place by the other methods than the collisions of ADAF-emitted MeV photons. As demonstrated in previous BH gap models, BH gaps are indeed capable of supplying such plasmas in a charge-starved magnetosphere via the cascade of the gap-emitted TeV photons into electron-positron pairs.

*3.3. Detectability of Gap Emissions*

As we have just seen, when the accretion rate becomes as small as:

$$\dot{m} < \dot{m}_{up} = 2.2\times 10^{-4} M_9^{-1/7} \tag{8}$$

the gap is 'switched on' and gamma rays are emitted by the gap-accelerated lepton. Let us examine if we can detect such gamma-rays. First, to constrain the range of accretion rate in which a BH gap can be activated, we plot the upper limit accretion rate, $\dot{m}_{up} = \dot{m}_{up}(M)$ (Equation (8)) as the thick solid line in Figure 1. Above this upper limit, the magnetosphere becomes force-free (blue shaded region). Thus, only under this thick solid line, BH gaps can be formed (white region). The dotted lines denote constant-$L_{BZ}$ lines as labeled.

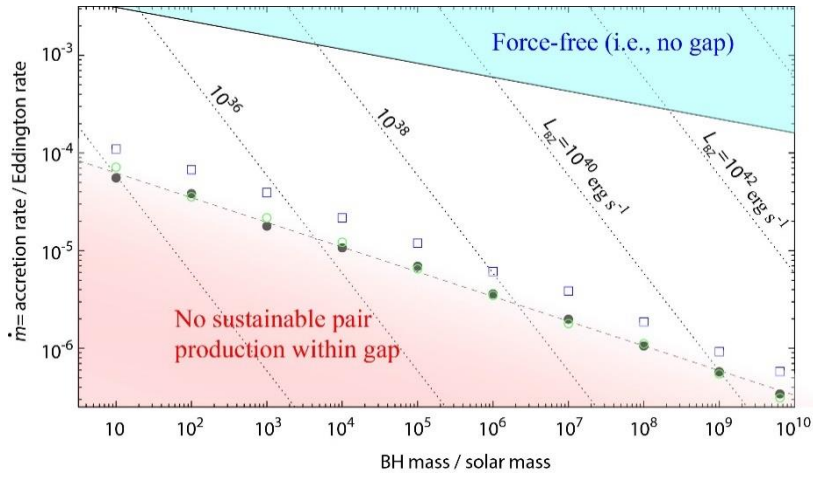

**Figure 1.** Gap formation region (white region) on the two-dimensional parameter space, (*M*, $\dot{m}$). The thick solid line shows the upper limit of the dimensionless accretion rate, $\dot{m}$, above which the copious pair production prevents the formation of a gap. The filled and open circles denote the lower limits of $\dot{m}$ (Section 5.4) for a stationary gap to exist in the case of *a*/*M* = 0.9 and 0.5, respectively. The thin dashed line shows a linear fit of the filled circles. The dotted lines show the Blandford–Znajek (BZ) power (Equation (4)) for an extremely rotating case, *a* = *M*. From [62].

Second, to examine the upper bound of the gap luminosity, we substitute $B = B_{eq}$ into Equation (4) to obtain the BZ power,

$$L_{BZ} = 1.7 \times 10^{46} \left(\frac{a}{M}\right)^2 \dot{m} M_9 \text{ erg s}^{-1}. \tag{9}$$

Putting *a* = *M*, we obtain the dotted lines in Figure 1.

Third, substituting $\dot{m} = \dot{m}_{up}$ into Equation (9), we obtain the maximum gap luminosity (as the cross section of the solid and dotted lines),

$$L_{BZ} = 3.7 \times 10^{42} \left(\frac{a}{M}\right)^2 M_9^{6/7} \text{ erg s}^{-1}. \tag{10}$$

Assuming that 100% of this power is converted into radiation, we obtain the upper limit of the gap flux at Earth, $F_{BZ} = L_{BZ}/4\pi d^2$, where *d* is the distance to the BH. Thus, for supermassive BHs, we obtain the following maximum flux,

$$F_{BZ} = 3.0 \times 10^{-10} \left(\frac{a}{M}\right)^2 M_9^{6/7} \left(\frac{d}{10\,\text{Mpc}}\right)^{-2} \text{ ergs s}^{-1}\text{ cm}^{-2}. \tag{11}$$

It follows that the IC component, which appears in VHE, may be detectable with ground-based IACTs. To quantify the actual flux as a function of the accretion rate, we must solve the gap electrodynamics. We will describe this issue in Sections 5–7.

*3.4. Criticism of the Stationary Black Hole (BH) Gap Model*

Levinson and Segev revisited the one-dimensional stationary BH gap model and found that the gap solutions are allowed only for small electric currents created within the gap [77]. They fully incorporated the GR effects in their basic equations, and solved the set of 1D Poisson equation, lepton Botzmann equations under mono-energetic approximation, and the radiative transfer equation, approximating all the IC photons gain the initial lepton kinetic energy, assuming a homogeneous and isotropic soft photon field (emitted by an ADAF). In their one-dimensional treatment, they found that $E_\parallel$ is proportional to $w^2$, and the potential drop, $V_{gap}$, is proportional to $w^3$ (because the Poisson equation is a second-order differential equation), as Figure 2a indicates. They imposed a boundary condition such that the created charge density matches the local GJ charge density at the outer boundary; accordingly, $j_{cr}$ is proportional $w$ (because of the Tayler expansion of $\rho_{GJ}$ around the null surface). Thus, $E_\parallel$ is proportional to $j_{cr}^{\ 2}$, and $V_{gap}$ to $j_{cr}^{\ 3}$. As a result of this strong dependence of $E_\parallel$ and $V_{gap}$ on $j_{cr}$, they found that stationary solutions can be found only for smaller current densities ($j_{LS} < 0.15$ in their definition) (Figure 2b), where $j_{LS}$ denotes the dimensionless current introduced by Levinson and Segev,

$$j_{LS} \equiv \frac{J_{cr}/2}{\Omega_F B_H \cos\theta/(2\pi)} = \frac{j_{cr}}{2\cos\theta}\left(\frac{r_H}{r}\right)^2 \qquad (12)$$

On the other hand, in the present review, we introduce the electric current per magnetic flux (5.4),

$$j_{cr} \equiv \frac{J_{cr}}{\omega_H B/(2\pi)}, \qquad (13)$$

Where $J_{cr}$ denotes the actual electric current. Thus, at $r \approx 1.4 r_H$ and $\theta \approx 0°$ for $\Omega_F \approx 0.5\omega_H$, $j_{LS} = 0.15$ corresponds to the conserved current of $j_{cr} \approx 0.3$ in our present notation (§5.4). Note that $j_{cr}$ is conserved along the individual magnetic flux tube.

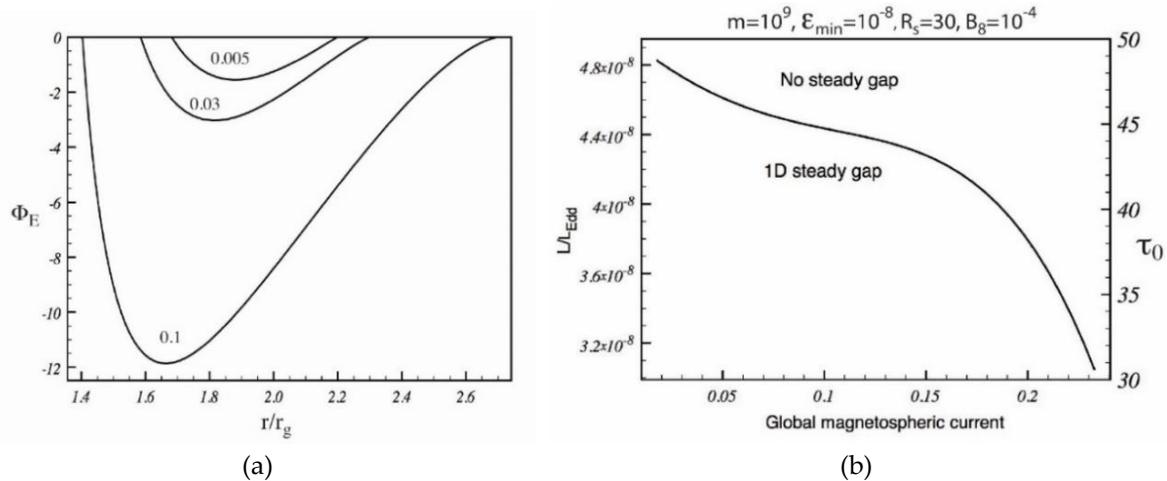

(a)　　　　　　　　　　　　　　　　　(b)

**Figure 2.** Stationary gap solutions for a billion solar-mass black hole (BH) for a split-monopole magnetic field line. (**a**) Renormalized acceleration electric field as a function of the Boyer–Lindquist radial coordinate. The three solid curves correspond to the solutions obtained for $j_{LS}$ = 0.005, 0.03, and 0.1 (Equation (12)). The peak of the non-corotational electric field, $\Phi_E$, is proportional to the square of the created current. (**b**) Maximum Eddington rate (i.e., maximum gap luminosity normalized by the Eddington luminosity) that a stationary gap can attain for a fix magnetospheric current, $j_{LS}$ (abscissa). Adapted with permission from [77].

It is worth noting, however, that the 2D screening effect of $E_\parallel$ becomes important when $w$ becomes comparable to or greater than the trans-field (e.g., meridional) thickness, $D_\perp$. In the case of pulsar OGs, $D_\perp$ becomes much less than the light cylinder radius, $c/\Omega_F$, for young pulsars like the Crab pulsar, and comparable to $c/\Omega_F$ for middle-aged pulsars like the Geminga pulsar (see §2 for references). For BHs, $D_\perp \sim r_g = M$ holds. Thus, in the present case, when $w \sim M$, $E_\parallel$ saturates and suppressed when $j_{cr} > 0.3$ (or $j_{LS} > 0.15$ in the notation of [78]). In this section, we present the result of a pulsar case, whose electrodynamics is essentially the same as BHs. In Figure 3, we show the pulsar's $E_\parallel$ solved for three discrete gap currents, $j_{cr} = 0.01$ (dashed curve), 0.03 (solid curve), and 0.0592 (dotted curve). As the gap width approaches $D_\perp = 3 \times 10^8$ cm, $E_\parallel$ is substantially suppressed by the 2D screening effect. As a result, pulsar OG solutions exist in a wide range of $j_{cr}$, namely $0 < j_{cr} < 1$. In the same way, for BHs, a stationary gap solution exists in a wide, $0 < j_{cr} < 1$, by virtue of the 2D screening effect, as will be demonstrated in §6.5. In another word, the difficulty of stationary BH gap model pointed out by [77], can be solved if we consider the 2D electrodynamic structure.

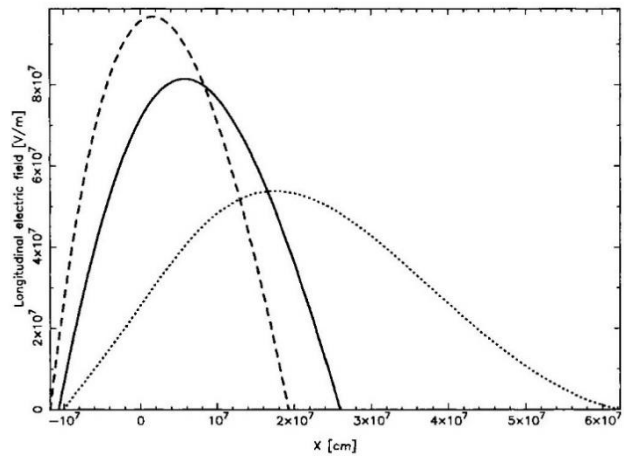

**Figure 3.** Acceleration electric field in a pulsar outer gap. The dashed, solid, and dotted curves correspond to the case of $j_{cr}$ = 0.01, 0.03, and 0.0592. From [78].

*3.5. Background Space-Time Geometry*

To examine the gap electrodynamics further quantitatively, let us describe the space time around a rotating BH. Since the self-gravity of the accreting plasmas and the electromagnetic field is negligible compared to the BH's gravitational field, the spacetime around a rotating BH is well described by the Kerr metric [79]. In the Boyer–Lindquist coordinate [80], it becomes:

$$ds^2 = g_{tt}dt^2 + 2g_{t\varphi}dtd\varphi + g_{rr}dr^2 + g_{\theta\theta}d\theta^2 + g_{\varphi\varphi}d\varphi^2 \tag{14}$$

where,

$$g_{tt} = -\frac{\Delta - a^2 \sin^2\theta}{\Sigma}, \quad g_{t\varphi} = -\frac{2Mar\sin^2\theta}{\Sigma}, \quad g_{\varphi\varphi} = \frac{A\sin^2\theta}{\Sigma}, \quad g_{rr} = \frac{\Sigma}{\Delta}, \quad g_{\theta\theta} = \Sigma, \tag{15}$$

and,

$$\Delta = r^2 - 2Mr + a^2, \quad \Sigma = r^2 + a^2 \cos^2\theta, \quad A = (r^2 + a^2)^2 - \Delta a^2 \sin^2\theta. \tag{16}$$

At the event horizon, $\Delta$ vanishes; thus, $r_H = r_g + \sqrt{r_g^2 - a^2}$ gives the horizon radius. The Schwarzschild radius is given by $2M$. If the BH is maximally rotating, $a \to M$ gives $r_H \to M$; that is, the horizon radius becomes half of the Schwarzschild radius. The space-time frame dragging angular frequency is given by:

$$\omega \equiv -\frac{g_{t\varphi}}{g_{\varphi\varphi}} = \frac{2Mar}{A}. \tag{17}$$

which decreases as $\omega \to 2Ma/r^3$ away from the BH, $r \gg M$. At the event horizon, $r = r_H$, on the other hand, becomes the BH's spin angular frequency,

$$\omega_H \equiv \frac{a}{2Mr_H}. \tag{18}$$

which does not depend on $\theta$, as expected. To lower an index of a tensor, we can use the following relations,

$$g^{tt} = -\frac{g_{\varphi\varphi}}{\rho_w^2}, \quad g^{\varphi\varphi} = -\frac{g_{tt}}{\rho_w^2}, \quad g^{t\varphi} = \frac{g_{t\varphi}}{\rho_w^2}, \quad g^{rr} = \frac{1}{g_{rr}}, \quad g^{\theta\theta} = \frac{1}{g_{\theta\theta}}, \tag{19}$$

where $\rho_w^2 \equiv g_{t\varphi}^2 - g_{tt}g_{\varphi\varphi} = \Delta\sin^2\theta$ vanishes at the horizon. We must choose appropriate vector and tensor components that are well-behaved at the horizon, as briefly discussed below in Equation (30).

*3.6. Non-Stationary BH Gap Model*

The aforementioned BH gap scenarios, however, are based on the stationary assumption. Thus, Levinson and Cerutti [81] examined the lepton acceleration and gamma-ray emission within a gap around a rotating BH, performing one-dimensional particle-in-cell (PIC) simulations. From the homogeneous part of the Maxwell equations (i.e., the Faraday law), one obtains:

$$\partial_t F_{r\theta} + \partial_r F_{\theta t} + \partial_\theta F_{tr} = 0, \tag{20}$$

$$\partial_t F_{\theta\varphi} + \partial_\theta F_{\varphi t} + \partial_\varphi F_{t\theta} = 0, \tag{21}$$

$$\partial_t F_{\varphi r} + \partial_\varphi F_{rt} + \partial_r F_{t\varphi} = 0, \tag{22}$$

where the electro-magnetic field strength tensor is related to the scalar and vector potentials by $F_{\mu\nu} = \partial_\mu A_\nu - \partial_\nu A_\mu$. They then investigated one-dimensional disturbances along the radial direction, setting $\partial_\theta = \partial_\varphi = 0$. Thus, they obtained:

$$\partial_t F_{r\theta} = -\partial_r F_{\theta t}, \tag{23}$$

$$\partial_t F_{\theta\varphi} = 0, \tag{24}$$

$$\partial_t F_{\varphi r} = -\partial_r F_{t\varphi}, \tag{25}$$

It follows from $\partial_t F_{\theta\varphi} = 0$ that the radial component of the magnetic field should be time-independent. As an example, they assumed a split-monopole solution, fixing the magnetic flux function to be $A_\varphi(r,\theta) = C(1-\cos\theta)$, where $C$ is a constant. This solution can be obtained around a slowly rotating BH if we give the toroidal current by $J^\varphi = Cr^{-4}$ on the equatorial plane [13,26]. With this solution of $A_\varphi$, the poloidal components of the magnetic field becomes time-independent and radial, because one obtains $F_{\theta\varphi} = \partial_\theta A_\varphi = C\sin\theta$ and $F_{\varphi r} = -\partial_r A_\varphi = 0$. Accordingly, we obtain $F_{\varphi t} = \partial_\varphi A_t - \partial_t A_\varphi = 0$; thus, the toroidal electric field vanishes. The poloidal components of the electric field are solved from the inhomogeneous part of the Maxwell equations (i.e., Amprere's law),

$$\nabla_\mu F^{\nu\mu} = \frac{1}{\sqrt{-g}} \partial_\mu \left( \sqrt{-g} F^{\nu\mu} \right) = 4\pi J^\nu, \tag{26}$$

where $J^\nu$ refers to the electric four current. Putting $\nu = r$ and, and noting that

$$\frac{\partial F^{rt}}{\partial t} = g^{rr} \left( g^{tt} \partial_t F_{rt} + g^{t\varphi} \partial_t F_{r\varphi} \right) = -\frac{A}{\Sigma^2} \left( \frac{\partial F_{rt}}{\partial t} + \omega \frac{\partial F_{r\varphi}}{\partial t} \right), \tag{27}$$

and,

$$\frac{\partial F^{\theta t}}{\partial t} = g^{\theta\theta} \left( g^{tt} \partial_t F_{\theta t} + g^{t\varphi} \partial_t F_{\theta\varphi} \right) = -\frac{A}{\Delta \Sigma^2} \left( \frac{\partial F_{\theta t}}{\partial t} + \omega \frac{\partial F_{\theta\varphi}}{\partial t} \right), \tag{28}$$

holds, we obtain:

$$\partial_t F_{rt} = \frac{\Sigma^2}{A} \partial_\theta \left( \ln \sqrt{-g} \right) F^{r\theta} - \frac{\Sigma^2}{A} \cdot 4\pi J^r, \tag{29}$$

and,

$$\partial_t F_{\theta t} = -\frac{\Delta \Sigma^2}{A} \partial_r \left( \ln \sqrt{-g} \right) F^{r\theta} - \frac{\Delta \Sigma^2}{A} \partial_r F^{r\theta} - \frac{\Delta \Sigma^2}{A} \cdot 4\pi J^\theta, \tag{30}$$

where $F_{r\theta} = g_{rr} g_{\theta\theta} F^{r\theta} = (\Sigma^2/\Delta) F^{r\theta}$; also, $\partial_\theta = \partial_\varphi = 0$ and $\partial_t F_{\theta\varphi} = \partial_t F_{\varphi r} = 0$ are used. Thus, they solve the time evolution of $F^{r\theta}$, $F_{rt}$, and $F_{\theta t}$, all of which are well-behaved at the horizon, from Equations (24), (29) and (30). The electric current density $J^\nu$ in Ampere's law is constructed from the actual motion of the charged particles in the PIC scheme.

By this one-dimensional, general-relativistic (GR) PIC simulation scheme, they demonstrated that $E_\parallel$ is efficiently screened out by the discharge of created pairs within the duration that is comparable to the dynamical time scale, $r_g/c = GM/c^3 = M$, that the residual $E_\parallel$ leads to a strong flare of VHE curvature radiation, which is followed by a state of self-sustained, rapid plasma oscillations, and that the resultant quasi-stationary gamma-ray luminosity attains only about $10^{-5}$ of the BZ power; that is, the time-averaged luminosity becomes less than the stationary analysis. See also Levinson et al., who semi-analytically found longitudinal oscillations in their two beam model [82].

Figure 4 exhibits the snapshots at four discrete simulation times. The upper panel shows the variation of the pair and photon densities. Pair density increases due to pair production from zero and saturates at about 10 times GJ value within $t = 5$ $GM/c^3 = 5$ $M$, that is, within a few dynamical time scales, where we may define the dynamical time scale to be $2$ $GM/c^3 = 2$ $M$. Accordingly, $E_\parallel$ is nearly screened out. The sporadic pair creation leads to rapid spatial and temporal oscillations of $E_\parallel$ (middle panel) and $J^r$ (bottom panel). Above nearly 10 dynamical time scales, $t > 20$ $M$, pair and photon distribution function attain quasi-stationary state.

Figure 5 exhibits the light curve evolution. It follows that the gamma-ray luminosity approaches the BZ power during the initial spike but decays to the terminal value as $E_\parallel$ is screened out. It is possible to argue that strong, rapid flares should be produced every time a magnetospheric gap is restored. The flaring episode is then followed by quiescent emission with a luminosity, $L_\gamma \sim 10^{-5} L_{BZ}$.

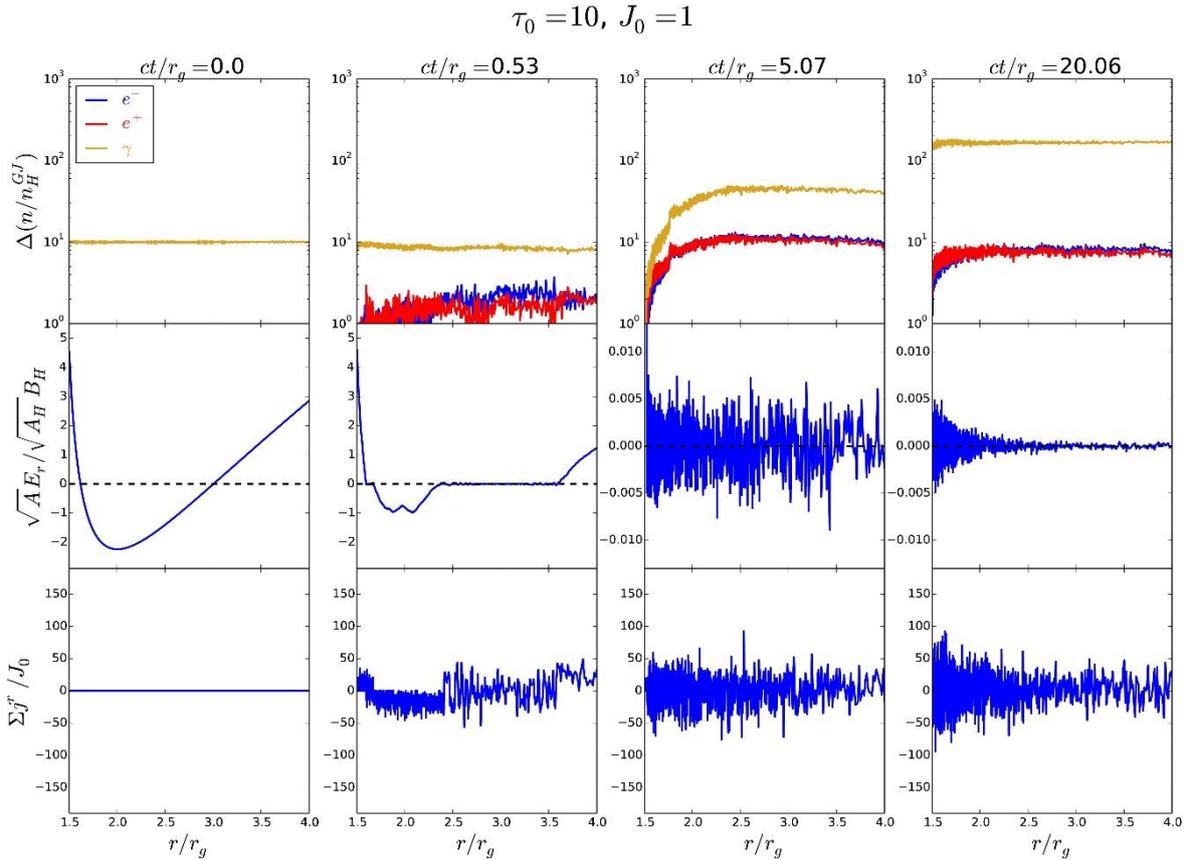

**Figure 4.** Snapshots from a typical simulation of spark gap dynamics. Shown are the pair and photon densities (upper panels), electric flux (middle panels), and radial electric current, $\Sigma J^r$, normalized by the global magnetospheric current $J_0$. The leftmost panels delineate the initial state, at *t = 0*. The rightmost panels show the relaxed state, following the prompt discharge. We note the scale change on the vertical axis in the middle panels. Adapted with permission from [81].

More recently, Chen et al. performed 1D PIC simulations on the BH gap model [83]. They assumed a Minkowski spacetime in the Maxwell equations, lepton equation of motion, and radiative transfer equation, except for the GJ charge density, whose GR expression is essential for the formation of a null surface near the event horizon, whereas full GR expressions were implemented in all the basic equations in [81]. Cheng et al. adopted a Newtonian split-monopole solution [84], and solved the time evolution of $E_\parallel = E_r$, $E_\theta$, and $B_\varphi$. It is found that a BH gap opens quasi-periodically and the screening of $E_\parallel$ creates oscillations. Eventually, when the created pairs escape across the light surfaces, the charge-starved magnetosphere recur in the formation of a gap.

Figure 6 shows the third cycle of such a gap development. The middle panel in the top indicates that $E_\parallel$ arises around the null surface, which is shown in the third panels as the crossing of the GR charge density (green curve in the orange band) and *0* (horizontal green line). However, the gap can also appear at different positions, as the right two panels in the top row show. The gap (shown in the top row) appears in the low multiplicity region, as indicated in the second row. Note that the green curve in the bottom panels show the spectra of the photons that materialize within the simulation box; thus, they do not show the gamma-ray spectra to be observed.

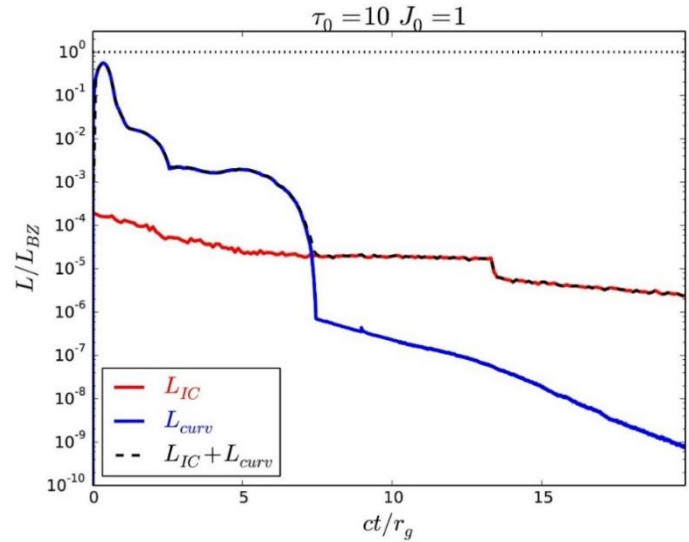

**Figure 5.** Gamma-ray light curve produced by the gap discharge. The red line corresponds to inverse-Compton (IC) emission, the blue line to curvature emission, and the black dashed line to the sum of both components. Adapted with permission from [81].

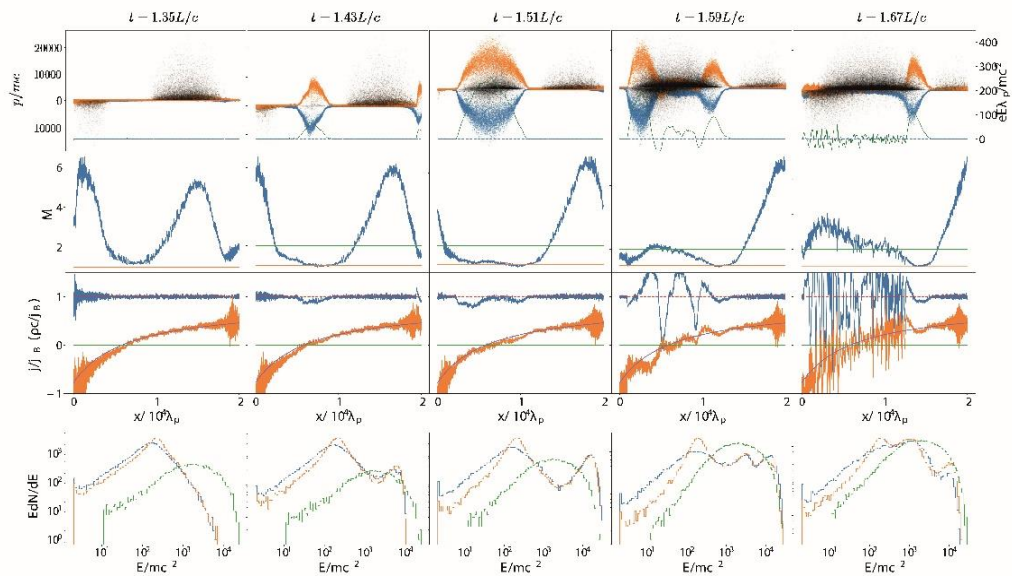

**Figure 6.** Time evolution of the gap. From left to right are snapshots at labeled times, where L is the size of the box. The panels from top to bottom are: (1) phase space plots for electrons (blue), positrons (orange), and photons (black). The green line is electric field and its scale is on the right. (2) Pair multiplicity *M*. Orange and green lines mark *M* = 1 and 2, respectively. (3) Current j (blue) and charge density (orange) and their background values. (4) Spectrum of electrons (blue), positrons (orange), and photons (green). Adapted with permission from [83].

In short, time-dependent, one-dimensional GR PIC simulations revealed that gaps are unsteady but tend to a quasi-steady state or a quasi-periodical state. The former, quasi-steady solution shows that the time-averaged gap gamma-ray luminosity is of an order of magnitude smaller than stationary analysis because of the efficient screening [81]. Nevertheless, in the present review we focus on the stationary BH gap scenario in order to elucidate the detailed gap electrodynamics, which will be also important to understand the time-dependent phenomena.

## 4. Advection-Dominated Accretion Flow

Before moving onto the gap electrodynamics, let us briefly describe the radiatively inefficient accretion flow in the lower latitudes (i.e., near the equator), because it is essential to quantify the photon–photon pair production and IC scatterings taking place inside and outside the gap in the higher latitudes (i.e., in the polar funnel). Since we are concerned with the soft photon field near the BH, we do not consider the structure of accretion flows away from the BH, focusing on an ADAF, a representative model of radiatively inefficient accretion flow. We consider that plasma accretion takes place near the rotational equator with a certain vertical thickness (e.g., [20,85,86]). Such an accretion cannot penetrate into the higher latitudes, that is, in the polar funnel because the centrifugal-force barrier prevents plasma accretion toward the rotation axis, and because the timescale for magnetic Rayleigh–Taylor instability or turbulent diffusion is long compared to the dynamical timescale of accretion. In this evacuated funnel, the poloidal magnetic field lines resemble a split monopole in a time-averaged sense [22]. The lower-latitude accretion emits MeV photons into the higher latitudes, supplying electron–positron pairs in the funnel. If the pair density in the funnel becomes less than $N_{GJ}$, $E_\parallel$ arises. Thus, a plasma accretion in the lower altitudes and a gap formation in the higher altitudes are compatible in a BH magnetosphere.

Since the gap luminosity increases with decreasing accretion rate [61–63,67,68], we consider a dimensionless accretion rate that is much small compared to unity. Such a low accretion status corresponds to the quiescence or low-hard state for X-ray binaries, and low luminosity active galactic nuclei for super-massive BHs. At such a low accretion rate, the equatorial accretion flow becomes optically thin for Bremsstrahlung absorption and radiatively inefficient because of the weak Coulomb interaction between the ions and electrons. Accordingly, the ions store the heat within the flow itself and accretes onto the BH without losing the thermal energy as radiation. This radiatively inefficient flow can be described by an ADAF [28,29,87-92], and provides the target soft photons for the IC-scattering and the photon-absorption processes in the polar funnel. Thus, to tabulate the redistribution functions for these two processes, we compute the specific intensity of the ADAF-emitted photons. For this purpose, we adopt the analytical self-similar ADAF spectrum presented in [90]. The spectrum includes the contribution of the synchrotron, IC, and Bremsstrahlung processes. These three cooling mechanisms balance with the heating due to the viscosity and the energy transport form ions, and determine the temperature of the electrons in an ADAF to be around $T_e \sim 10^9$ K. In radio wavelength, the ADAF radiation field is dominated by the synchrotron component whose peak frequency, $\nu_{c,syn}$, varies with the accretion rate as $\nu_{c,syn} \propto \dot{m}^{1/2}$ In the X-ray wavelength, the Bremsstrahlung component dominates the ADAF flux at such a low $\dot{m}$. In soft gamma-ray wavelength, this component cuts off around the energy $h\nu \sim kT_e$. These MeV photons (with energies slightly below $kT_e$ collide each other to materialize as seed electrons and positrons that initiate a pair-production cascade within the gap. If the accretion rate is as low as $\dot{m} < \dot{m}_{up}(M,a) < 10^{-2.5}$, the seed pair density becomes less than the GJ value [59], thereby leading to an occurrence of a vacuum gap in the funnel.

It is noteworthy, however, that at low accretion rate, namely $\dot{m} < 10^{-3}$, less efficient thermalization may lead to a hybrid thermal-nonthermal energy distribution. Observationally, it is indeed known that non-thermal electrons are needed to reproduce the quiescent low-frequency radio emission in Sgr A* [93–95] and other low-luminosity AGNs [96]. Theoretically, the direct heating of electrons (i.e., not via Coulomb collisions with ions) is proposed via several mechanisms, such as magnetic reconnection [97–101], MHD turbulence [98,102–105], or dissipation of pressure anisotropy in a collisionless plasma [106]. At the present stage, there is no consensus on the theory of electron heating yet. In the present review, we neglect such direct heating of electrons and assume that only less than 1% of the ions thermal energy is transferred to electrons. However, if more gravitational energy is converted into electrons' heat by direct heating (i.e., if the fraction of electron heating becomes » 0.01), the increased ADAF soft photon field would significantly reduce the gap luminosity at a fixed accretion rate. We should keep this point in mind when we construct a BH gap model, irrespective of whether the gap is stationary or non-stationary. We will present the input ADAF spectrum when we show the predicted gamma-ray spectrum of the gap emissions.

## 5. Stationary, Two-Dimensional Gap Electrodynamics

Next, we formulate the BH gap, which will be formed when $\dot{m} < \dot{m}_{\text{up}}$.

### 5.1. Poisson Equation for the Non-Corotational Potential

In a stationary and axisymmetric spacetime, Gauss's law for the electrostatic potential gives

$$\nabla_\mu F^{t\mu} = \frac{1}{\sqrt{-g}} \partial_\mu \left[ \frac{\sqrt{-g}}{\rho_w^2} g^{\mu\nu} (-g_{\varphi\varphi} F_{t\nu} + g_{t\varphi} F_{\varphi\nu}) \right] = 4\pi\rho, \tag{31}$$

where $\nabla$ denotes the covariant derivative, $\sqrt{-g} = \sqrt{g_{rr} g_{\theta\theta} \rho_w^2} = \Sigma \sin\theta$, $\rho_w^2 = \Delta \sin^2\theta$, and denotes the real charge density; the Greek indices run over $t$, $r$, $\theta$, $\varphi$. The electro-magnetic field strength tensor $F_{\mu\nu}$ is defined in section 3.6.

In the present formalism, we assume that the electromagnetic fields depend on $t$ and $\varphi$ only through $\varphi - \Omega_F t$, where $\Omega_F$ denotes the angular frequency of the magnetic field rotation. Imposing the ideal MHD condition (i.e., imposing a very large magnetic Reynolds number in the Ohm's law, which leads to a vanishing electric field in the co-moving frame of the fluid), we obtain $F_{rt} = \Omega_F F_{\varphi r}$ and $F_{\theta t} = -\Omega_F F_{\theta\varphi}$. Under this assumption of stationarity, $F_{\mu\nu} = F_{\mu\nu}(r, \theta, \varphi - \Omega_F t)$, we can introduce the non-corotational potential such that:

$$F_{\mu t} + \Omega_F F_{\mu\varphi} = \partial_\mu \Phi(r, \theta, \varphi - \Omega_F t), \tag{32}$$

If $F_{At} + \Omega_F F_{A\varphi}$ vanishes for $A=r$ and $\theta$, $\Omega_F$ is conserved along the magnetic field line. However, in a lepton accelerator, $F_{At} + \Omega_F F_{A\varphi}$ deviates from 0 and the magnetic field lines do not rigidly rotate. The deviation from rigid rotation is expressed by how $\Phi$ deviates from a constant value at each place in the magnetosphere.

We adopt the static observer whose four velocity is given by the Killing vector, $\xi^\mu = (1,0,0,0)$. Then the electromagnetic fields have the following components in the Boyer–Lindquist coordinates, $E_r = F_{rt}$, $E_\theta = F_{\theta t}$, $E_\varphi = F_{\varphi t}$, and [107,108]

$$B^r = -\frac{g_{tt} + g_{t\varphi}\Omega_F}{\sqrt{-g}} F_{\theta\varphi}, \quad B^\theta = -\frac{g_{tt} + g_{t\varphi}\Omega_F}{\sqrt{-g}} F_{\varphi r}, \quad B_\varphi = \sqrt{-g} F^{r\theta}. \tag{33}$$

The acceleration electric field is given by $E_\parallel = \mathbf{E} \cdot \mathbf{B}/B = (F_{it} + \Omega_F F_{i\varphi})B^i/B = (\mathbf{B}/B) \cdot (\nabla\Phi)$, where $B \equiv |\mathbf{B}|$. Note that $(E_\parallel B)^2 = (\mathbf{E} \cdot \mathbf{B})^2 = \det(F)$ is a frame-independent scalar.

Substituting Equation (30) into (29), we obtain the Poisson equation for the non-corotational potential,

$$\frac{1}{\sqrt{-g}} \partial_\mu \left[ \frac{\sqrt{-g}}{\rho_w^2} g^{\mu\nu} g_{\varphi\varphi} \partial_\nu \Phi \right] = 4\pi(\rho - \rho_{\text{GJ}}), \tag{34}$$

where the general-relativistic Goldreich–Julian (GJ) charge density is defined by

$$\rho_{\text{GJ}} \equiv \frac{1}{4\pi\sqrt{-g}} \partial_\mu \left[ \frac{\sqrt{-g}}{\rho_w^2} g^{\mu\nu} g_{\varphi\varphi} (\Omega_F - \omega) F_{\varphi\nu} \right], \tag{35}$$

where $\omega \equiv -g_{t\varphi}/g_{\varphi\varphi}$ denotes the frame dragging angular frequency. Away from the horizon, $r \gg M$, Equation (35) reduces to the standard, special-relativistic expression of the GJ charge density,

$$\rho_{GJ} = -\frac{\mathbf{\Omega} \cdot \mathbf{B}}{2\pi c} + \frac{(\mathbf{\Omega} \times \mathbf{r}) \cdot (\nabla \times \mathbf{B})}{4\pi c} \quad (36)$$

Not only $F_{\mu\nu}$ and $\Phi$ but also $\rho$ may depend on $t$ and $\varphi$ through $\varphi - \Omega_F t$. In this case, equation (34) gives "stationary" gap solution in the "co-rotational" frame, in the sense that $\Phi$ and $\rho$ are a function of $r$, $\theta$, and $\varphi - \Omega_F t$. Note that such stationary solutions are valid not only between the two light surfaces where $k_0 \equiv -g_{tt} - 2g_{t\varphi}\Omega_F - g_{\varphi\varphi}\Omega_F^2 > 0$, but also inside the inner light surface and outside the outer light surface where $k_0 < 0$ [109-111].

*5.2. The Null-Charge Surface*

Equation (34) shows that $E_\parallel$ is exerted along $\mathbf{B}$ if $\rho$ deviates from $\rho_{GJ}$. In a stationary gap, the derivative of $E_\parallel$ along $\mathbf{B}$ should change sign at the outer and the inner boundaries, where the outer boundary is located at larger distance from the hole, while the inner boundary is located near the hole. Thus, a vacuum gap (with $\rho \approx 0$) is located around the null-charge surface, on which $\rho_{GJ}$ changes sign. If the gap becomes non-vacuum (i.e., $|\rho|$ is slight smaller than $|\rho_{GJ}|$), the gap position is essentially unchanged (6.6). Thus, the null-charge surface becomes the natural place for a gap to arise, in the same way as the pulsar outer-gap model (for an analytic argument of the gap position, see 2 of [112]).

In the case of BHs, the null-charge surface appears near the surface on which $\omega$ coincides with $\Omega_F$ [52]. Since $\omega$ matches $\Omega_F$ only near the horizon, the gap inevitably appears near the horizon, irrespective of the BH mass. In Figure 7, we plot the distribution of the null surface as the thick red solid curve. In the left panel, we assume that the magnetic field is split-monopole, adopting $A_\varphi = C(1-\cos\theta)$ as the magnetic flux function [13], where $C$ is a constant. In the right panel, on the other hand, we assume a parabolic magnetic field line with $A_\varphi = (C/2)r(1-\cos\theta)$ on the poloidal plane (i.e., $r$–$\theta$ plane). We adopt $\Omega_F = 0.3\omega_H$ for both cases [20,113]. It follows that the null surface distributes nearly spherically, irrespective of the poloidal magnetic field configuration. This is because its position is essentially determined by the condition $\omega(r,\theta) = \Omega_F(A_\varphi)$, because $\omega$ has weak dependence on $\theta$, and because we assume $\Omega_F$ is constant for $A_\varphi$ for simplicity. If $\Omega_F$ decreases (or increases) toward the polar region, the null surface shape becomes prolate (or oblate). The field angular frequency $\Omega_F$ is, indeed, deeply related to the accretion conditions. For instance, in order to get adequate jet efficiency from the magnetic field in the funnel above the BH for a radio-loud active galactic nuclei (AGNs), one needs significant flux compression from the lateral boundary imposed by the accretion flow and corona [114]. Although it is not a self-consistent solution, we adopt a constant $\Omega_F$ in the present review for simplicity.

Let us quickly take a look at the case of slower and faster rotations. Figure 8 shows the GJ charge density and the null-charge surface distribution when the poloidal magnetic field line is radial (i.e., split-monopole); the BH mass and spin are the same as Figure 7. The left panel corresponds to the case of $\Omega_F = 0.15\omega_H$, while the right panel $\Omega_F = 0.60\omega_H$. It follows that the null-charge surface approaches the horizon for as $\Omega_F$ increases. In what follows, we adopt $\Omega_F = 0.5\omega_H$ unless explicitly mentioned, because the BZ power maximizes at this magnetic-field rotation.

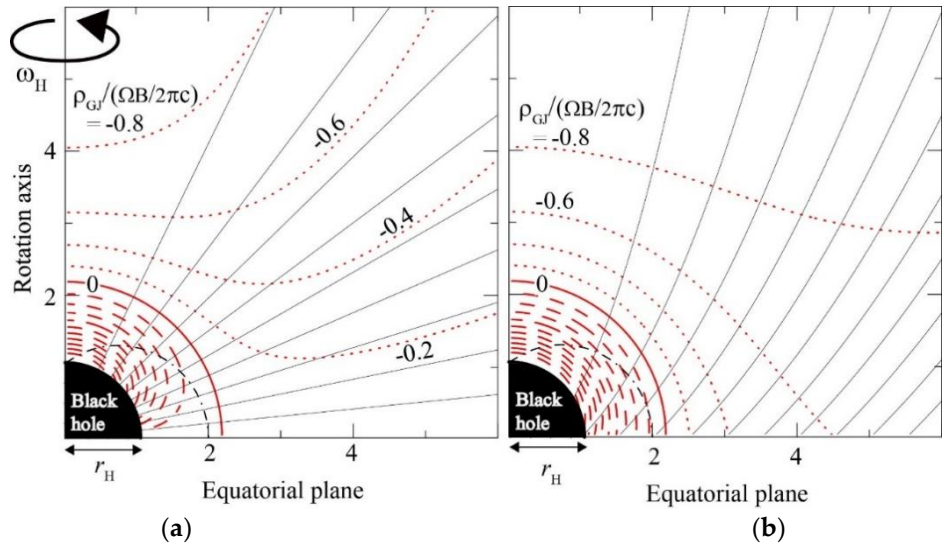

**Figure 7.** Distribution of the null surface (red thick solid curve) on the poloidal plane in the Boyer–Lindquist coordinates. The axes are in $r_g = GM/c^2 = M$ unit. The black hole (filled black region in the lower left corner) rotates rapidly with spin parameter $a = 0.998\,M$ around the ordinate. The contours of the dimensionless Goldreich–Julian charge density are plotted with the red dashed curves (for positive values) and the red dotted ones (for negative values as labeled). The black dash-dotted curve denotes the static limit, within which the rotational energy of the hole is stored. The black solid curves denote the magnetic field lines. (**a**) The case of a split-monopole magnetic field. (**b**) The case of a parabolic field. From [61].

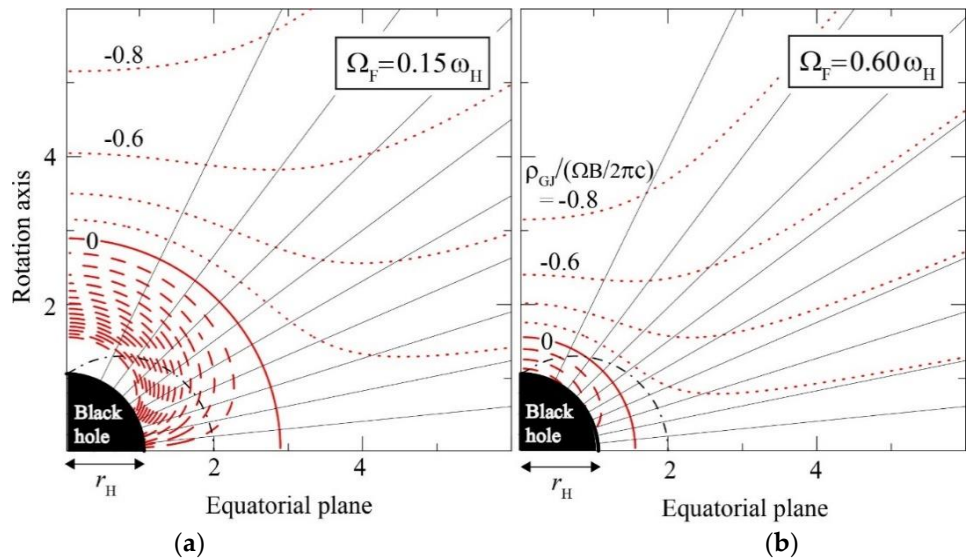

**Figure 8.** Similar to the left panel of Figure 7: both panels show the dimensionless Goldreich–Julian charge density (red dashed, solid, and dotted contours) for a split-monopole magnetic field (black solid lines). (**a**) The case of a slower field-line rotation, $\Omega_F = 0.15\omega_H$. (**b**) The case of a faster rotation, $\Omega_F = 0.60\omega_H$. From [61].

### 5.3. The Stagnation Surface

In a stationary and axisymmetric BH magnetosphere, an MHD fluid flows from a greater $k_0$ region to a smaller $k_0$ one. It follows that both the inflows and outflows start from the two-dimensional surface on which $k_0$ maximizes along the poloidal magnetic field line. Thus, putting $k_0' = 0$, we obtain the position of the stagnation surface (thick green solid curve in Figure 9), where the prime denotes the derivative along the poloidal magnetic field line. The condition $k_0' = 0$ is equivalent to imposing a balance among the gravitational, centrifugal, and Lorentz forces on the poloidal plane. In Figure 9, we plot the contours of $k_0$ for split-monopole and parabolic poloidal magnetic field lines.

It is clear that the stagnation surface is located at $r < 5\,M$ in the lower latitudes but at $r > 5\,M$ in the higher latitudes. As the poloidal field geometry changes from radial to parabolic, the stagnation surface moderately moves away from the rotation axis in the higher latitudes. This analytical argument of the position of the stagnation surface [115], was confirmed by general relativistic (GR) MHD simulations [60]. In these numerical works, the stagnation surface is time-dependent but stably located at 5–10 $M$ with a prolate shape, as depicted in Figure 9. The two-dimensional surfaces on which $k_0$ vanishes, are called the outer and inner light surfaces (thick red solid curves in Figure 9). Outside the outer light surface (or inside the inner light surface), plasma particles must flow outwards (or inwards).

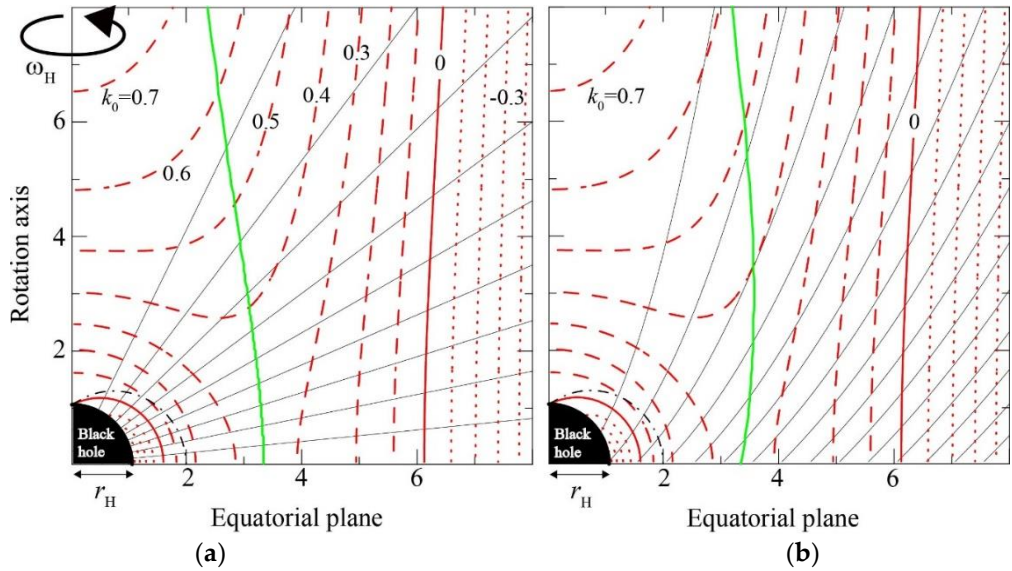

**Figure 9.** Distribution of the stagnation surface (thick green solid curve) on the poloidal plane. Similar to Figure 7, but equi-$k_0$ potential curves are depicted with the dashed curves (for positive values) and dotted ones (for negative values). The thick, red solid curves denote the outer and inner light surfaces where $k_0$ vanishes. (**a**) The case of a split-monopole magnetic field. (**b**) The case of a parabolic field. From [61].

*5.4. The Gap Position*

To examine the Poisson equation (34), it is worth noting that the distribution of $\rho_{\mathrm{GJ}}$ has relatively small dependence on the magnetic field geometry near the horizon, because its distribution is essentially governed by the space-time dragging effect, as indicated by the red solid, dashed, and dotted contours in Figure 7 for split-monopole and parabolic fields. As a result, the gap solution little depends on the magnetic field line configuration, particularly in the higher latitudes. In what follows, we thus assume a split-monopole field and adopt a constant $\Omega_F$ for $A_\varphi$ for simplicity, as described in the left panels of Figures 7 and 9.

In a pulsar magnetosphere, the gap position is located around the null-charge surface if there is no leptonic injection across either of the inner and outer boundaries, and that the gap position shifts outwards (or inwards) when leptons are injected into the gap across the inner (or the outer) boundary [116,117]. Since the gap electrodynamics is essentially the same between pulsar outer gaps and BH gaps, we can expect that the same conclusion holds in the case of BH magnetospheres.

In a BH magnetosphere, a gap indeed distributes around the null-charge surface, as will be shown in sections 6.3 and 7.2.

*5.5. Advection-Dominated Accretion Flow (ADAF)-Emitted Soft Photon Field in the Magnetosphere*

To quantify the gap electrodynamics, we need to compute the pair creation rate. To this end, we must tabulate the specific intensity of the soft photons at each position in the polar funnel. In this section, we therefore consider how the soft photons are emitted in an ADAF and propagate around a rotating BH. We assume that the soft-photon field is axisymmetric with respect to the BH rotation axis. When the mass accretion rate is much smaller than the Eddington rate, the accreting plasmas form an ADAF with a certain thickness in the equatorial region. For simplicity, in this paper we approximate that such plasmas rotate around the BH with the GR Keplerian angular velocity,

$$\Omega_K \equiv \pm \frac{1}{M} \frac{1}{(r/M)^{3/2} \pm a/M} \tag{37}$$

and that their motion is dominated by this rotation. That is, we neglect the motion of the soft-photon-emitting plasmas on the poloidal plane, $(r, \theta)$, for simplicity.

Let us introduce the local rest frame (LRF) of such rotating plasmas. The orthonormal tetrad of the LRF is given by [63],

$$\mathbf{e}_{(\hat{t})}^{\mathrm{LRF}} = \left(\frac{dt}{d\tau}\right)^{\mathrm{LRF}} \left(\partial_t + \Omega_K \partial_\varphi\right), \tag{38}$$

$$\mathbf{e}_{(\hat{\varphi})}^{\mathrm{LRF}} = \frac{g_{t\varphi} + g_{\varphi\varphi}\Omega_K}{\rho_w \sqrt{D}} \partial_t - \frac{g_{tt} + g_{t\varphi}\Omega_K}{\rho_w \sqrt{D}} \partial_\varphi, \tag{39}$$

$$\mathbf{e}_{(\hat{r})}^{\mathrm{LRF}} = \sqrt{g^{rr}} \partial_r, \qquad \mathbf{e}_{(\hat{\theta})}^{\mathrm{LRF}} = \sqrt{g^{\theta\theta}} \partial_\theta, \tag{40}$$

where,

$$D \equiv -g_{tt} - 2g_{t\varphi}\Omega_K - g_{\varphi\varphi}\Omega_K^2, \tag{41}$$

and the redshift factor between the LRF and the distant static observer becomes:

$$\left(\frac{d\tau}{dt}\right)^{\mathrm{LRF}} = \sqrt{D}. \tag{42}$$

Note that the GR Keplerian angular frequency on the equator, $\Omega_K$, is different from the magnetic-field angular frequency, $\Omega_F$. Thus, $D$ is not $k_0$. Note also that equation (42) is the reciprocal of what appears in equation (38). For further details on how to compute the specific intensity of the ADAF soft photon field, see [63].

Between the distant static observer (i.e., us) and the LRF, the photon energy changes by the redshift factor,

$$\frac{\omega_\infty}{\omega_{\mathrm{LRF}}} = \frac{\mathbf{e}_{(t)}^{\infty} \cdot \mathbf{k}}{\mathbf{e}_{(t)}^{\mathrm{LRF}} \cdot \mathbf{k}} = \left(\frac{dt}{d\tau}\right)^{\mathrm{LRF}} \frac{-\hbar\omega_\infty}{(\mathbf{e}_{(t)} + \beta^\varphi \mathbf{e}_{(\varphi)}) \cdot \mathbf{k}} = \left(\frac{dt}{d\tau}\right)^{\mathrm{LRF}} \frac{1}{1 - \beta^\varphi \lambda}, \tag{43}$$

where $\lambda \equiv k_\varphi / \hbar\omega_\infty$ denotes the ratio between the photon angular momentum $k_\varphi$ and energy $-k_t = \hbar\omega_\infty$, both of which are conserved along the ray.

We tabulate the soft photon specific intensity at each position in the magnetosphere by the ray-tracing method. The dispersion relation $k^\mu k_\mu = 0$ gives the Hamiltonian,

$$H = -k_t = -\frac{g_{t\varphi}}{g_{\varphi\varphi}} k_\varphi \pm \frac{\rho_w}{g_{\varphi\varphi}} \sqrt{k_\varphi^2 + g_{\varphi\varphi}(g^{rr} k_r^2 + g^{\theta\theta} k_\theta^2)}, \tag{44}$$

Thus, the Hamilton–Jacobi relation gives $dr/dt$, $d\theta/dt$, $dk_r/dt$, and $dk_\theta/dt$ in terms of $(r,\theta)$ and $(k_r, k_\theta)$ for a fixed $(-k_t, k_\varphi)$, which enables us to ray-trace the soft photons in the Kerr space time [63].

In what follows, we compute the collision frequencies of two-photon pair creation and IC scatterings in the frame of a zero-angular-momentum observer (ZAMO), which rotates with the same angular frequency as the space-time dragging frequency, $\omega = -g_{t\varphi}/g_{\varphi\varphi}$. Putting $\beta^\varphi = \omega$, we obtain the lapse (section 3.3 of [63]):

$$\frac{d\tau}{dt} = \alpha = \frac{\rho_w}{\sqrt{g_{\varphi\varphi}}} = \sqrt{\frac{\Delta\Sigma}{A}}, \qquad (45)$$

Between the distant static observer and ZAMO, the photon energy changes by the redshift factor,

$$\frac{\omega_\infty}{\omega_{\mathrm{ZAMO}}} = \frac{\mathbf{e}_{(t)}^\infty \cdot \mathbf{k}}{\mathbf{e}_{(t)}^{\mathrm{ZAMO}} \cdot \mathbf{k}} = \frac{\alpha}{1 - \beta^\varphi \lambda}, \qquad (46)$$

Combining equations (46) and (58), we find that the photon energy changes from the LRF (in the disk) to ZAMO (in the magnetosphere including the polar funnel) by the factor:

$$g_s \equiv \frac{\omega_{\mathrm{ZAMO}}}{\omega_{\mathrm{LRF}}} = \alpha^{-1}\left(\frac{d\tau}{dt}\right)^{\mathrm{LRF}} \frac{1-\omega\lambda}{1-\Omega_K\lambda}, \qquad (47)$$

By the ray-tracing technique described above (see also [118]), we can now compute the specific intensity of the soft photons emitted from the ADAF at each place in the magnetosphere in the ZAMO frame of reference. Integrating this ZAMO-measured specific intensity over the propagation solid angle in each directional bin at each point in ZAMO, we obtain the soft photon flux at each point in each direction. Finally, we use this flux to compute the IC optical depth and the photon–photon collision optical depth in ZAMO.

Let us define $(dF_s/dE_s)_0$ to be the differential soft photon flux that would be obtained if the specific intensity were homogeneous and isotropic in a Minkowski space time within the radius *6M*, and if the total soft photon luminosity were given by radial flux at *6M* multiplied by $4\pi(6M)^2$. Then, the photon differential number flux at each place can be expressed as $dF_s/dE_s = f_{ADAF}(r,\theta)(dF_s/dE_s)_0$, where the flux correction factor can be computed by summing up the ray-traced photons at each place by the method described above. If the photon specific intensity were homogeneous and isotropic at $r < 6M$ in a flat spacetime, we would have $f_{ADAF}(r,\theta) = 1$ at any $r$ and $\theta$ within the radius *6M*.

Figure 10 shows the flux correction factor, $f_{ADAF}(r,\theta)$. The top left, top right, bottom left and bottom right panels show the results for colatitudes $\theta = 0°, 15°, 30°, 45°$, respectively. The black solid, red dashed, green dot-dashed, blue dotted, and cyan triple-dot-dashed curves denote $f_{ADAF}$ at $r = 2\,M$, $4\,M$, $6\,M$, $15\,M$, and $30\,M$, respectively. The results do not depend on the BH mass, because the radius is normalized by the gravitational radius, $r_g = M$. In each panel, the abscissa indicates the photon momentum projected along the radial outward direction; thus, +1 means radially outward propagation while −1 does radially inward one. It follows that the photon intensity decreases with decreasing $\theta$, because most of the photons have positive angular momentum $\lambda$ and hence find it difficult to approach the rotation axis, $\theta = 0$. The solid curves (at $r = 2M$) in each panel show that the radiation field becomes predominantly inward near the horizon due to the causality, where the horizon is located at $r_H = 1.435M$ when $a = 0.9M$. However, at larger radius *r*, the radiation field becomes outwardly unidirectional and its flux decreases by the $r^{-2}$ law, as the blue (at $r = 15M$) and cyan (at $r = 30M$) curves indicate. This is because the ADAF photons are emitted only within $r_{out} = 10M$ in the present consideration.

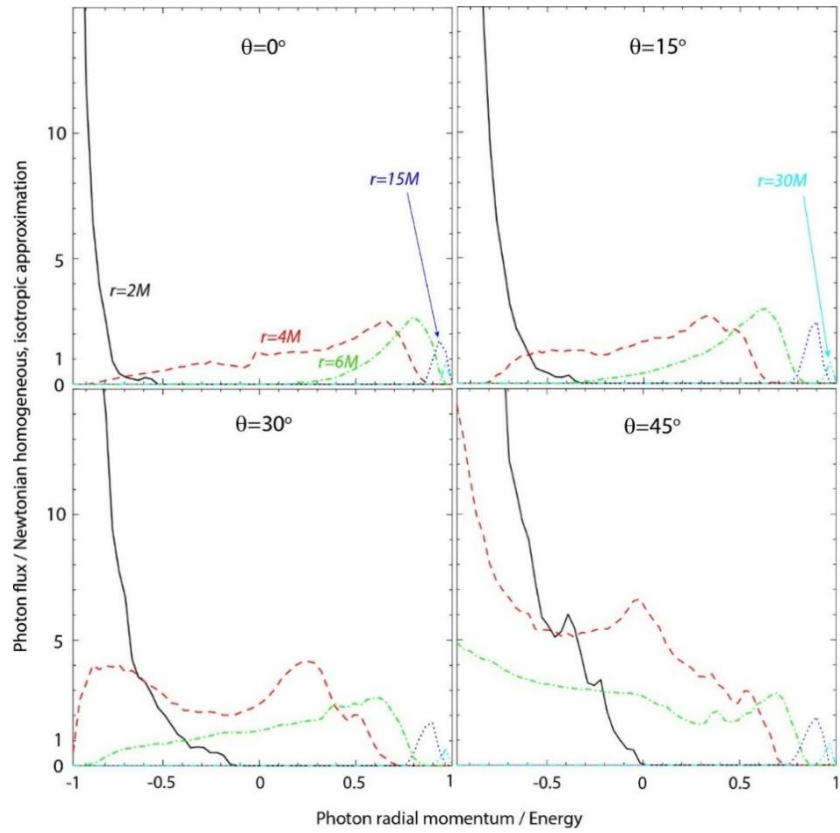

**Figure 10.** Advection-dominated accretion flow (ADAF) photon flux correction factor, $f_{ADAF}$, as a function of the cosine of the propagation direction, $\cos\theta_\gamma^{ZAMO}$, with respect to the radially outward direction, at five discrete radial positions at three latitudes, as labeled. BH spin is assumed to be *0.9 M*. The black solid, red dashed, green dot-dashed, blue dotted, and cyan triple-dot-dashed curves denote $f_{ADAF}$ at *r* = 2 *M*, 4 *M*, 6 *M*, 15 *M*, and 30 *M*, respectively. In each panel, the abscissa changes from $\cos\theta_\gamma^{ZAMO} = -1$ to +1; −1 corresponds to the radially inward direction, whereas +1 does radially outward direction in the zero-angular-momentum observer (ZAMO). From [63].

*5.6. The Poisson Equation in the Tortoise Coordinate*

To solve the radial dependence of *Φ* in the Poisson equation (34), we introduce the following dimensionless tortoise coordinate,

$$\frac{d\eta_*}{dr} = \frac{r^2 + a^2}{\Delta}\frac{1}{M}, \tag{48}$$

In this coordinate, the horizon corresponds to the "inward infinity," $\eta_* \to -\infty$. In this paper, we set $\eta_* = r/M$ at *r* = 25*M*. Note that the function $\eta_* = \eta_*(r/M)$ does not depend on $\theta$. For a graph of $\eta_* = \eta_*(r/M)$, see Figure 2 of [62].

Using the tortoise coordinate, the Poisson equation (34) can be cast into the form

$$\frac{A}{\Sigma^2}\left(\frac{r^2+a^2}{\Delta}\right)^2\frac{\partial^2\Phi}{\partial\eta_*^2} + \left[\frac{A}{\Sigma^2}\frac{2r(a^2-Mr)}{\Delta^2} + \frac{r^2+a^2}{\Sigma\Delta}\frac{\partial}{\partial r}\left(\frac{A}{\Sigma}\right)\right]\frac{\partial\Phi}{\partial\eta_*}$$
$$+ \frac{A}{\Sigma^2\Delta}\frac{\partial^2\Phi}{\partial\theta^2} + \frac{1}{\Sigma\Delta\sin\theta}\frac{\partial}{\partial\theta}\left(\frac{A\sin\theta}{\Sigma}\right)\frac{\partial\Phi}{\partial\theta} = \frac{B(\eta_*,\theta)}{B_H}\left[\int(n_+ - n_-)d\gamma - n_{GJ}\right], \tag{49}$$

The $\csc\theta$ factor in the $\partial_\theta$ term can be eliminated if we change the variable such as $z = \cos\theta$.

The dimensionless lepton distribution functions per magnetic flux tube are defined by

$$n_\pm(\eta_*,\theta;\gamma) \equiv \frac{2\pi c e}{\omega_H B(\eta_*,\theta)} N_\pm(\eta_*,\theta;\gamma), \tag{50}$$

whereas the dimensionless GJ number density is defined by:

$$n_{GJ}(\eta_*,\theta) \equiv \frac{2\pi c}{\omega_H B(\eta_*,\theta)} \rho_{GJ}(\eta_*,\theta), \tag{51}$$

$N_\pm$ denotes the distribution function of positrons and electrons in ordinary definition. We have introduced the dimensionless non-corotational potential,

$$\Phi(\eta_*,\theta) \equiv \frac{c}{2M^2 \omega_H B_H} \Phi(r,\theta), \tag{52}$$

For a radial poloidal magnetic field, $A_\varphi = A_\varphi(\theta)$, we can compute the acceleration electric field by

$$E_\parallel \equiv \frac{\partial \Phi}{\partial r} = \frac{2M\omega_H B}{c} \frac{r^2 + a^2}{\Delta} \frac{\partial \Phi}{\partial \eta_*}, \tag{53}$$

Without loss of any generality, we can assume $F_{\theta\varphi} > 0$ in the northern hemisphere. In this case, a negative $E_\parallel$ arises in the gap, which is consistent with the direction of the global current flow pattern.

*5.7. Boltzmann Equations for Electrons and Positrons*

Next, let us describe the lepton Boltzmann equations. Imposing a stationary condition, $\partial_t + \Omega_F \partial_\varphi = 0$, we obtain the Boltzmann equations,

$$\cos\chi \frac{\partial n_\pm}{\partial s} + \dot{p}\frac{\partial n_\pm}{\partial p} = \alpha\left(S_{IC,\pm} + S_{p,\pm}\right), \tag{54}$$

Where $\chi$ denotes the pitch angle (with $\chi = 0$ for out-going leptons and $\chi = \pi$ for in-coming ones), *s* and *p* show the position and dimensionless momentum along the magnetic field, $S_{IC}$ denotes the rate of positrons or electrons transferred into Lorentz factor $\gamma$ from another Lorentz factor via IC scatterings, $S_P$ refers to that which appeared into Lorentz factor $\gamma$ via photon–photon and magnetic pair production, and $\alpha$ refers to the lapse function, or equivalently, the gravitational redshift factor. In the Boyer-Lindquist coordinate, we obtain

$$\alpha = \frac{\rho_w}{\sqrt{g_{\varphi\varphi}}} = \sqrt{\frac{\Delta\Sigma}{A}}, \tag{55}$$

It follows that $\alpha = 0$ at the horizon and $\alpha = 1$ at infinity. We consider both photon–photon and magnetic pair production process; however, only the former contributes in all the cases we consider. The upper and lower sign of n, $S_{IC}$, and $S_P$ corresponds to the positrons (with charge $q = +e$) and electrons ($q = -e$), respectively. The dimensionless momentum is related to the Lorentz factor by $p \equiv |\mathbf{p}| = m_e c \sqrt{\gamma^2 - 1}$, where $m_e$ shows the electron rest mass. The left-hand side is in *dt* basis, where *t* denotes the proper time of a distant static observer. Thus, the lapse $\alpha$ is multiplied in the right-hand side, because both $S_{IC}$, and $S_P$ are evaluated in ZAMO. It is convenient to include the curvature emission as a friction term in the left-hand side. In this case, we obtain:

$$\dot{p} \equiv qE_{\parallel}\cos\chi - \frac{P_{SC}}{c}, \tag{56}$$

where the curvature radiation force is given by (e.g., [119]), $P_{SC}/c = 2e^2\gamma^4/(3R_c^2)$. The IC and pair-production redistribution functions are given in [50,63].

It is noteworthy that the charge conservation ensures that the dimensionless total current density (per magnetic flux tube),

$$j_{tot}(s) = \int\left[-n_+(s,\gamma) - n_-(s,\gamma)\right]d\gamma, \tag{57}$$

is conserved along the flowline, where electrons migrate outwards while positron inwards. If we denote the created current density as $J_{cr}$, the injected current density across the inner and outer boundaries as $J_{in}$ and $J_{out}$, respectively, and the typical GJ value as $J_{GJ} = \omega_H B_H/2\pi$, we obtain $j_{tot} = j_{cr} + j_{in} + j_{out}$, where $j_{cr} \equiv J_{cr}/J_{GJ}$, $j_{in} \equiv J_{in}/J_{GJ}$, $j_{out} \equiv J_{out}/J_{GJ}$.

### 5.8. Radiative Transfer Equation

We assume that all photons are emitted with vanishing angular momenta and hence propagate on a constant-$\theta$ cone. Under this assumption of radial propagation, we obtain the radiative transfer equation:

$$\frac{dI_\omega}{dl} = -\alpha_\omega I_\omega + j_\omega, \tag{58}$$

where $dl = \sqrt{g_{rr}}dr$ refers to the distance interval along the ray in ZAMO, $\alpha_\omega$ and $j_\omega$ the absorption and emission coefficients evaluated in ZAMO, respectively. We consider only photon–photon collisions for absorption, pure curvature and IC processes for primary lepton emissions, and synchrotron and IC processes for the emissions by secondary and higher-generation pairs. For more details of the computation of absorption and emission, see Section 4.2 and 4.3 of [61] and Section 5.1.5 of [62].

### 5.9. Boundary Conditions

The elliptic type second-order partial differential Equation (49) is solved on the 2-D poloidal plane. We assume a reflection symmetry, $\partial_\theta\Phi = 0$, at $\theta = 0$. We assume that the polar funnel is bounded at a fixed colatitude, $\theta = \theta_{max}$ and that this lower-latitude boundary is equi-potential and put $\Phi = 0$ at $\theta = \theta_{max} = 60°$. Both the outer and inner boundaries are treated as free boundaries. At both inner and outer boundaries, $E_\parallel = -\partial\Phi/\partial r$ vanishes. To solve the Boltzmann equation (54), we need to specify the particle injection rate across the inner and outer boundaries. Since the magnetospheric current is to be constrained by a global condition including the distant dissipative region, we should treat $j_{cr}$, $j_{in}$ and $j_{out}$ as free parameters, when we focus on the local gap electrodynamics. For simplicity, we assume that there is no electron injection across the inner boundary and put $j_{in} = 0$ throughout this paper. In what follows, we examine stationary gap solutions for several representative values of $j_{cr}$ and $j_{out}$. The radiative-transfer Equation (58), a first-order ordinary differential, contains no photon injection across neither the outer nor the inner boundaries.

### 5.10. Gap Closure Condition

We compute the multiplicity (Eq.[41] of [61]) of primary electrons, $M_{out}$, and that of primary positrons, $M_{in}$, summing up all the created pairs by individual test particles and dividing the result by the number of test particles. With this modification, we apply the same closure condition that a stationary gap may be sustained, $M_{out}M_{in} = 1$.

## 6. Lepton Accelerator around Stellar-Mass Black Holes

Before we come to the main subject, the case of super-massive BHs, it is desirable to investigate stellar-mass BHs, because important gap electrodynamics have so far been elucidated in this case [67,68].

*6.1. Black Hole Accretion in a Gaseous Cloud*

When a BH moves in an ambient medium, accretion takes place onto the BH. Since the temperature is very low in a molecular cloud, the BH's velocity *V* becomes supersonic with respect to the gaseous cloud. For a uniform medium, particles within the impact parameter $r_B \approx GM/V^2$ will be captured by the BH with the hydrodynamical accretion rate [120,121] $\dot{M}_B = 4\pi\lambda_B (GM)^2 \rho V^{-3}$, where we obtain $\lambda_B = 1.12$ for an isothermal gas, and 0.25 for an adiabatic one; $\rho$ denotes the mass density of the gas. Thus, we obtain the dimensionless Bondi accretion rate:

$$\dot{m}_B = 5.39 \times 10^{-9} \lambda_B n_{H_2} \left(\frac{M}{10 M_\odot}\right) \left(\frac{\eta}{0.1}\right)^{-1} \left(\frac{V}{10^2 \text{ km s}^{-1}}\right)^{-3}, \qquad (59)$$

for a molecular hydrogen gas, where $n_{H_2}$ denotes the density of $H_2$ in the unit of cm$^{-3}$, $\eta \sim 0.1$ designates the radiation efficiency of the accretion flow. Representative values of $\dot{m}_B$ are plotted as the five straight lines in Figure 11a. Since the accreting gases have little angular momentum as a whole with respect to the BH, they form an accretion disk only within a radius that is much less than the Bondi radius $r_B$. Thus, we neglect the mass loss as a disk wind between $r_B$ and the inner-most region, and evaluate the accretion rate near the BH, $\dot{m}$ with $\dot{m}_B$. In section 6, we consider a 10-solar-mass BH, which is typical as a stellar-mass BH [122,123].

*6.2. Lepton Accelerator around Stellar-Mass Black Holes*

In a vacuum, rotating magnetosphere, an acceleration electric field, $E_\parallel$, inevitably arises. Accordingly, leptons (red arrows in Figure 11b) are accelerated into ultra-relativistic energies to emit high-energy gamma-rays (wavy line with middle wavelength) via the curvature process and VHE gamma-rays (wavy line with shortest wavelength) via the inverse-Compton (IC) scatterings of the soft photons (wavy line with longest wavelength) that were emitted from the ADAF. A fraction of such VHE photons collide with the ADAF soft photons to materialize as $e^\pm$ pairs, which partially screen the original $E_\parallel$ when they polarize. To compute the actual strength of $E_\parallel$, we solve the $e^\pm$ pair production cascade in a stationary and axisymmetric magnetosphere on the meridional plane $(r, \theta)$. The magnetic field lines are twisted into the counter-rotational direction due to the poloidal currents, and the curvature radius of such a toroidal field is assumed to be $r_g = M$ in the local reference frame.

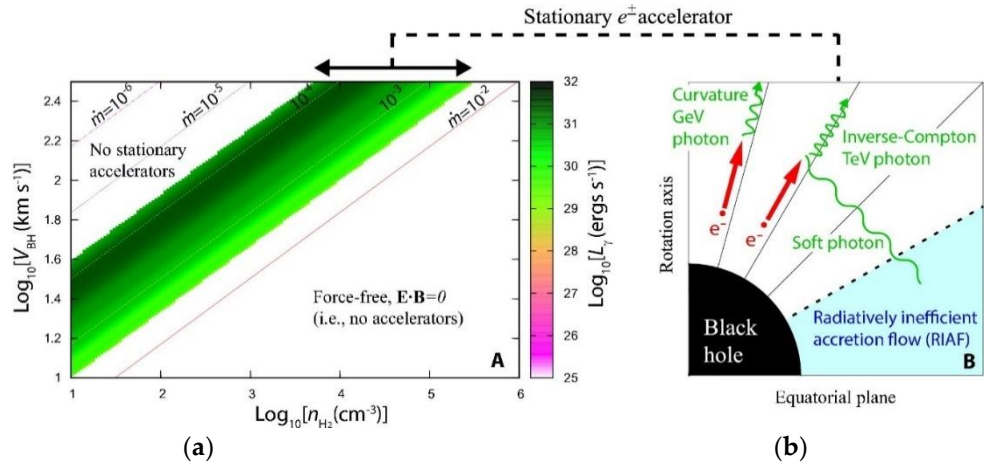

**Figure 11.** (**a**) Luminosity of BH lepton accelerators when a 10-solar-mass, extremely rotating (*a* = 0.99*M*) BH is moving with velocity *V* in a cloud whose molecule hydrogen density is $n_{H_2}$. The five straight lines correspond to the Bondi-Hoyle accretion rates, $10^{-2}$, $10^{-3}$, $10^{-4}$, $10^{-5}$, and $10^{-6}$, as labeled. In the lower-right white region, the gap vanishes due to efficient pair production. In the upper-left white region, stationary accelerators cannot be formed. Thus, stationary accelerators arise only in the green-black region. (**b**) Schematic figure (side view) of a BH magnetosphere. The polar funnel is assumed to be bounded from the ADAF, a kind of radiatively inefficient accretion flow (RIAF, cyan region) at colatitude *θ* = 60◦ (dashed line) from the rotation axis (i.e., the ordinate). From [67].

### 6.3. Results: Acceleration Electric Field

To estimate the maximum luminosity of a gap, we consider an extremely rotating case of *a* = 0.99 (Figure 4 of [68]). It follows from Figure 12a that $E_\parallel$ peaks along the rotation axis, because the magnetic fluxes concentrate towards the rotation axis as *a*→*M*. In what follows, we thus focus on the emission along the rotation axis, $\theta = 0°$. $E_\parallel$ decreases slowly outside the null surface in the same way as pulsar outer gaps (e.g., Figure 12 of [124]). This is due to the two-dimensional screening effect of $E_\parallel$.

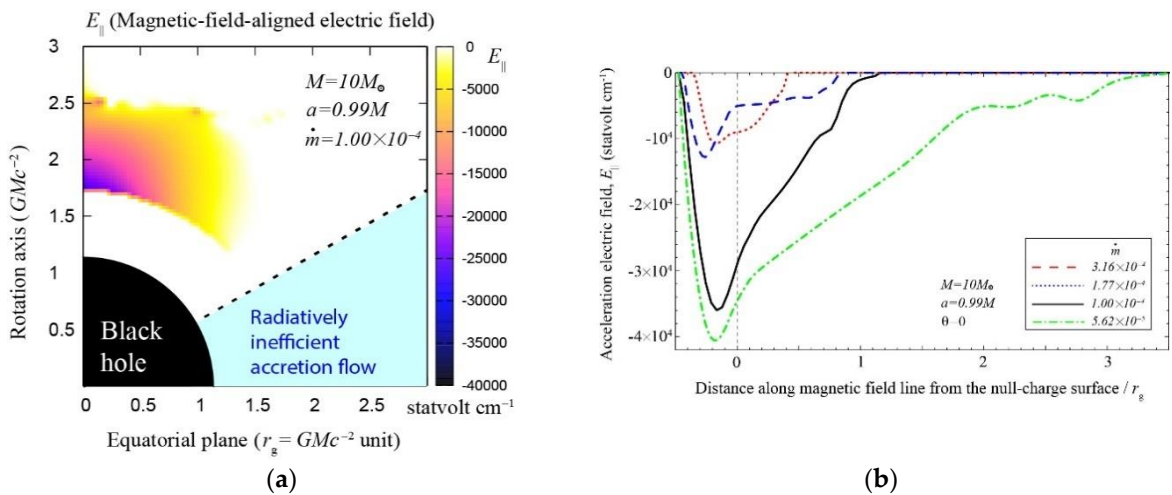

**Figure 12.** Acceleration electric field, $E_\parallel$, for *M* = 10*M*⊙, *a* = 0.99*M*, and split-monopole magnetic field lines rotating with angular frequency $\Omega_F = 0.5 \omega_H$. (**a**) Distribution on the poloidal plane; both axes are normalized by the gravitational radius, $r_g$. Strong *E*∥ appears only in the polar region. The mass accretion rate is $10^{-4}$ in the Eddington unit. The null-charge surface distributes nearly spherically at radis, *r* ≈ 1.73*M*. (**b**) $E_\parallel$ as a function of distance from the null surface (abscissa) along the rotation axis. The created current density is 70% of the Goldreich–Julian value, and the injected currents are set to be zero. The red dashed, blue dotted, green dash-

dotted, and black solid curves corresponds to $\dot{m} \equiv \dot{M}/\dot{M}_{Edd}$ = 3.16 × 10$^{-4}$, 1.77 × 10$^{-4}$, 1.00 × 10$^{-4}$, and 5.62 × 10$^{-5}$, respectively. The vertical dashed line shows the position of the null-charge surface. From [67,68].

In Figure 12b, we present $E_\parallel(s, \theta = 0°)$ at four dimensionless accretion rates (as indicated in the box). We find that the potential drop increases with decreasing $\dot{m}$. However, if the accretion further decreases as $\dot{m} < 10^{-4.25}$, there exists no stationary gap solutions. Below this lower bound accretion rate, the gap becomes inevitably non-stationary (section 8.1).

### 6.4. Results: Ultra-Relativistic Leptons

Since $E_\parallel < 0$, electrons (or positrons) are accelerated outward (or inward), the electrons' Lorentz factors, $\gamma$, gather around 3×10$^6$ due to the curvature-radiation drag (Figure 2 of [68]). At the same time, $\gamma$ distributes in lower values with a broad plateau in 6 × 10$^4$ < $\gamma$ < 2 × 10$^6$ due to the IC drag. Since the Klein–Nishina cross section increases with decreasing $\gamma$, such lower-energy electrons with 6 × 10$^4$ < $\gamma$ < 2 × 10$^6$ efficiently contribute to the VHE emission via IC scatterings [68].

### 6.5. Results: Gamma-Ray Spectra

Let us examine the gamma-ray spectra. In Figure 13a, we present the spectral energy distribution (SED) of the gap emission along five discrete viewing angles. It follows that the gap luminosity maximizes if we observe the BH almost face-on, $\theta < 15°$, and that the gap luminosity rapidly decreases at $\theta < 30°$.

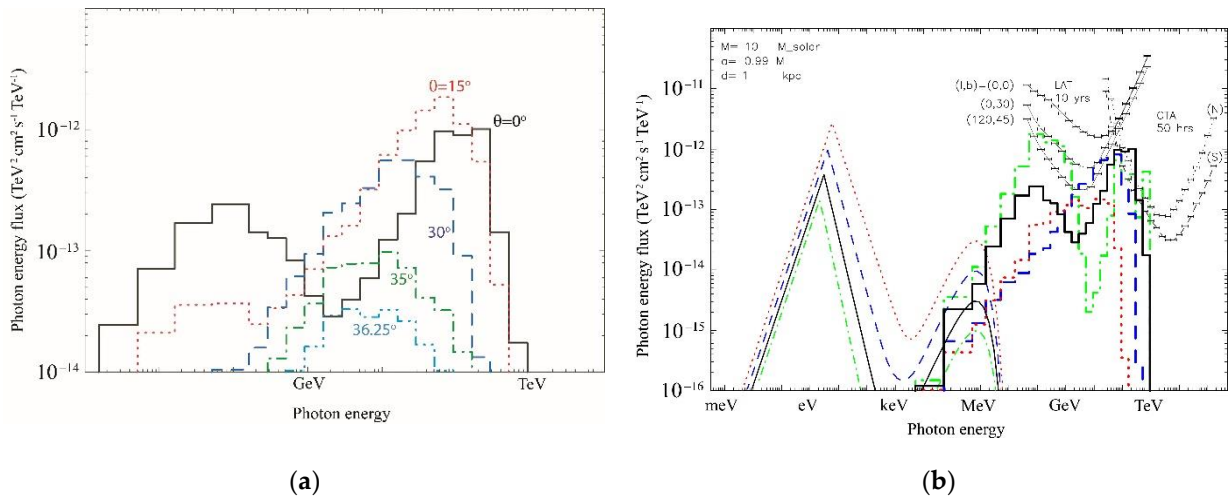

(**a**)　　　　　　　　　　　　　　　(**b**)

**Figure 13.** Spectral energy distribution (SED) of the gap-emitted photon for the BH whose mass and spin are *M* = 10 $M_\odot$ and *a = 0.99 M*. The created and injected current densities are $j_{cr} = 0.7$, $j_{in} = 0$, $j_{out} = 0$. (**a**) SEDs along five discrete viewing angles with respect to the rotation axis; the black solid, red dotted, blue dashed, green dot-dashed, and cyan triple-dot-dashed lines correspond to $\theta$ = 0°, 15°, 30°, 35°, and 36.25°, respectively. The accretion rate is $\dot{m}$ =1.00×10$^{-4}$. (**b**) SEDs at four discrete accretion rate; the red dotted, blue dashed, black solid, and green dot-dashed lines correspond to $\dot{m}$ =3.16 × 10$^{-4}$, 1.77 × 10$^{-4}$, 1.00 × 10$^{-4}$, and 5.62 × 10$^{-5}$, respectively. From [67,68].

In Figure 13b, we show the SEDs for *a* = 0.99*M* at $\dot{m}$ = 10$^{-3.50}$, 10$^{-3.75}$, 10$^{-4.00}$, and 10$^{-4.25}$ along $\theta = 0°$. It follows that the gap luminosity increases with decreasing $\dot{m}$. The VHE flux lies above the CTA detection limit (dashed and dotted curves on the top right), provided that the distance is within 1 kpc and we observe it nearly face on (i.e., $\theta \sim 0°$). Indeed, the VHE appears above the CTA detection limit if *a* > 0.90 *M* (Figure 4 [68]).

However, if the BH is moderately rotating as $a$ = 0.50 $M$, it is very difficult to detect its emission, unless it is located within 0.3 kpc [68]. It also follows that the spectrum has two peaks. The curvature photons are emitted by the gap-accelerated primary electrons and appear between 5 MeV and 0.5 GeV. The same electrons up-scatter the ambient soft photons, forming the second peak between 0.5 GeV and 5 GeV. The latter IC photons suffers absorption above 0.1 TeV (Figure 5 of [68]).

### 6.6. Results: Dependence on the Current Created in the Gap

Let us demonstrate that the gap solution exists in a wide range of the created electric current within the gap, $j_{cr} \equiv J_{cr}/J_{GJ}$, and that the resultant gamma-ray spectrum little depends on $j_{cr}$. In Figure 14, we show the solved $E_{\parallel}(s)$ (left panels) and SEDs (right panels) for three discrete $j_{cr}$'s: from the top, they correspond to $j_{cr}$ = 0.3, 0.5, and 0.9. The case of $j_{cr}$ = *0.7* is presented as Figure 13b. It is clear that the gap spectra modestly depend on the created current within the gap as long as the created current is sub-GJ. There exist no stationary gap solutions if $j_{cr} > 1$. Note that the maximum value of $|E_{\parallel}|$ saturates below 4.5 × 10$^5$ statvolt cm$^{-1}$ because of the 2D screening effect, which becomes important when the gap width become comparable to or greater than the transverse thickness ~$M$ (3.4).

It can also be shown that the SEDs depend little on the injected current across the outer boundary (Figure 7 of [68]). For further details, including the authors' comments to a criticism raised by [77] regarding the locations of stationary gaps and the stagnation surface, see section 6.2 of [68].

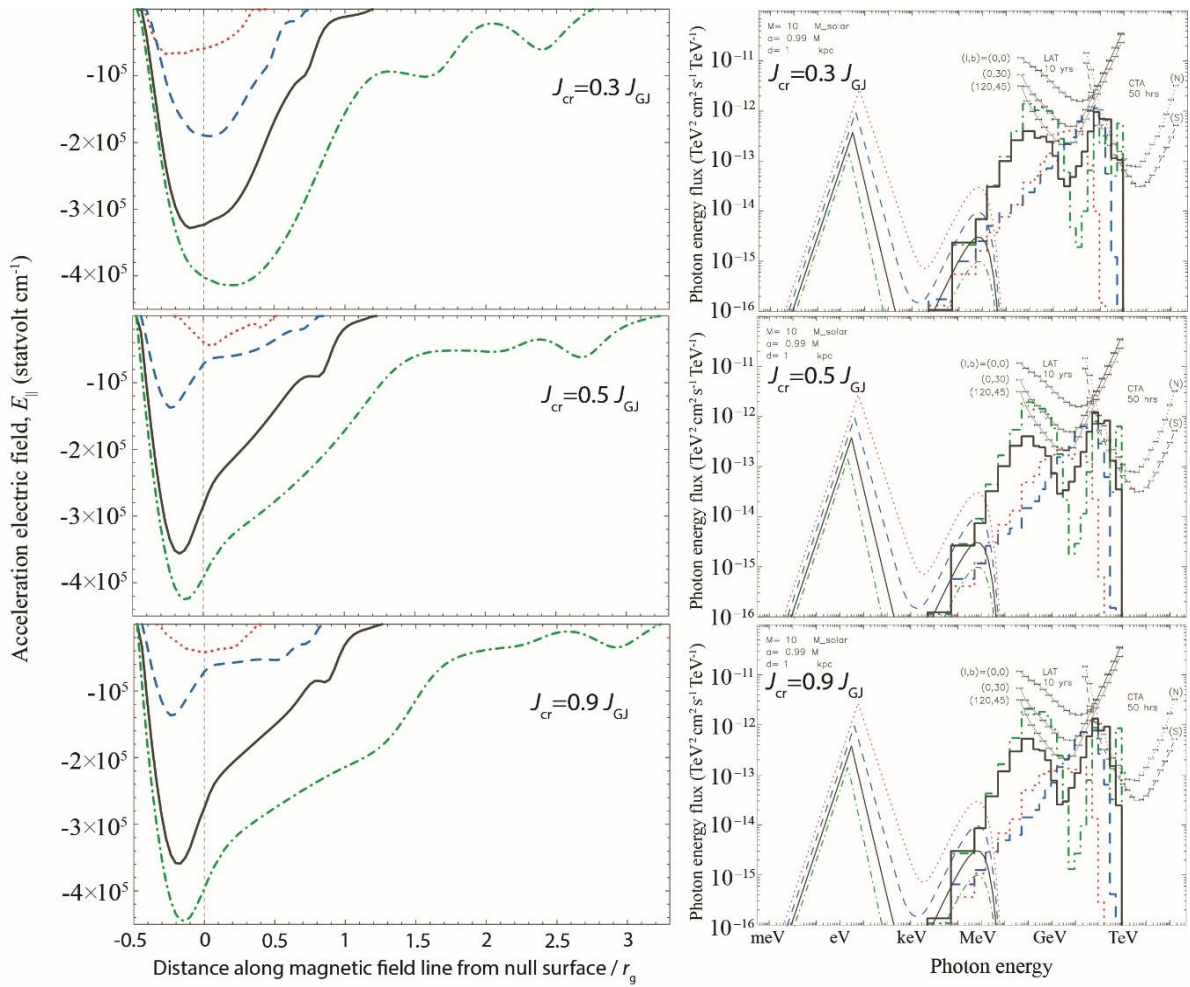

**Figure 14.** Dependence of the magnetic field-aligned electric fields $E_{\parallel}(s)$ (left panels) and the SEDs (right panels) on the created current within the gap. From the top, the created current density is 30%, 50%, and 90% of the GJ value. In all the three cases, the injected current density across the inner or outer boundaries is set to be zero. The curves corresponds to the accretion rate of $\dot{m}$ =3.16×10$^{-4}$ (red dotted), 1.77 × 10$^{-4}$ (blue dashed), 1.00 × 10$^{-4}$ (black solid), and 5.62 × 10$^{-5}$ (green dash-dotted). From [68].

## 7. Lepton Accelerator around Super-Massive Black Holes

In this section, we apply the BH gap model to supermassive BHs.

### 7.1. Very High-Energy Observations of Active Galaxies

The High-Energy Stereoscopic System (H.E.S.S.), the Major Atmospheric Gamma Imaging Cherenkov (MAGIC), and the Very Energetic Radiation Imaging Telescope Array System (VERITAS) have so far detected 219 TeV sources [125]. Among them, 70 sources have been identified as AGNs. Although most of them are blazer type, NGC 1275, 3C264, M87, and Cen A are classified as Fanaroff–Riley (FR) I radio galaxies. In addition, another radio galaxy, IC 310 appears to be of a transitional type between FR I and BL Lac. For these five non-blazer radio galaxies, their inner jet (on parsec or sub-parsec scales) are moderately misaligned with respect to the line of sight. Thus, their TeV emission from the jets will not be highly Doppler-boosted. Accordingly, the emission from the direct vicinity of the central supermassive BH, if it exists, may not be hidden by the strong jet emissioin. We are, therefore, motivated to study these non-blazer radio galaxies, comparing theoretical predictions with VHE observations. See Rieger and Levinson [126] for a comprehensive review on the VHE observations of these non-blazer radio galaxies, and associated theoretical arguments.

From NGC 1275, a TeV flare has been reported by MAGIC [127] with the shortest variation time scale of $\Delta t_{\text{var}}$ ~10 h. Since the light crossing time of the event horizon is about 3 × 10$^3$ s for this source, the dimension of the TeV emitting region should be less than 10 times Schwarzschild radii.

M87 exhibited rapidly varying VHE flares that are observed in 2008 and 2010 [128–131], where the shortest variation time scale was observed by all the three IACTs, and was found to be $\Delta t_{\text{var}}$ ~1 day [128]. Its horizon-crossing time scale, 4 × 10$^4$ s, indicates that the TeV emitting region should be less than a few Schwarzschild radii. Also, VHE photons were detected during the other periods, including the low-emission state [132–134]. Results of multi-wavelength observations of M87 were summarized in [135].

IC 310 showed very rapid variabilities whose shortest time scale attained $\Delta t_{\text{var}}$ ~5 m = 300 s, which was detected by MAGIC [77]. Compared with its horizon-crossing time scale of 3 × 10$^3$ s, we find that the TeV photons should be emitted from a sub-horizon length scale, presumably from the direct vicinity of the event horizon. For a comparison of representative emission models for such rapidly varying extragalactic sources in the TeV sky, see [136].

In the rest of this review, we consider the BH gap model as a possible explanation of such horizon-scale or sub-horizon-scale TeV flares observed from M87 and IC 310.

### 7.2. Results: Acceleration Electric Field

To investigate the VHE emission from the direct vicinity of a supermassive BH, we apply the BH gap model, and comapre the results obtained for $M$ = 10$^9$ M$_\odot$ and 10$^6$ M$_\odot$. Unless explicitly mentioned, we adopt $a = 0.90M$, $\Omega_F = 0.5\omega_H$, $j_{\text{tot}} = 0.7$, $j_{\text{in}} = j_{\text{out}} = 0$, and the curvature radius is $R_C = M$ (due to toroidal bending of the magnetic field lines). To solve the Poisson equation (49), we set the meridional boundary at 60°.

Let us begin with the $E_\parallel$ distribution on the poloidal plane ($\eta_*, \theta$), where $\eta_* \equiv r_*/M$ denotes the dimensionless tortoise coordinate (5.4). In Figure 15, we plot $E_\parallel$ (in statvolt cm$^{-1}$) for billion solar-mass BH (left panel) and for million solar-mass BH (right panel). For both panels, a common accretion rate, $\dot{m}$ =1.77×10$^{-5}$, is adopted. Near the lower-latitude boundary, $\theta$ =60°, a small $E_\parallel$ extends into higher altitudes, because a smaller $E_\parallel$ requires a greater width for the gap closure condition to be satisfied. This behavior is common with pulsar lower-altitude slot-gap models [137,138]. However, near the rotation axis, $\theta$ = 0°, stronger $E_\parallel$ arises, mainly because the polar region is far from the meridional boundary at $\theta$ = 60°, and partly because the magentic fluxes more or less concentrate toward the rotation axis even for a non-extreme rotator with *a* = 0.90*M*. The strong $E_\parallel$ distributes relatively widely in the polar region within $\theta \leq 15°$. $E_\parallel$ decreases with increasing BH mass, because the GJ charge density, $\rho_{GJ} \approx \Omega B / 2\pi c$, decreases with increasing BH mass, where $B = B_{eq}$.

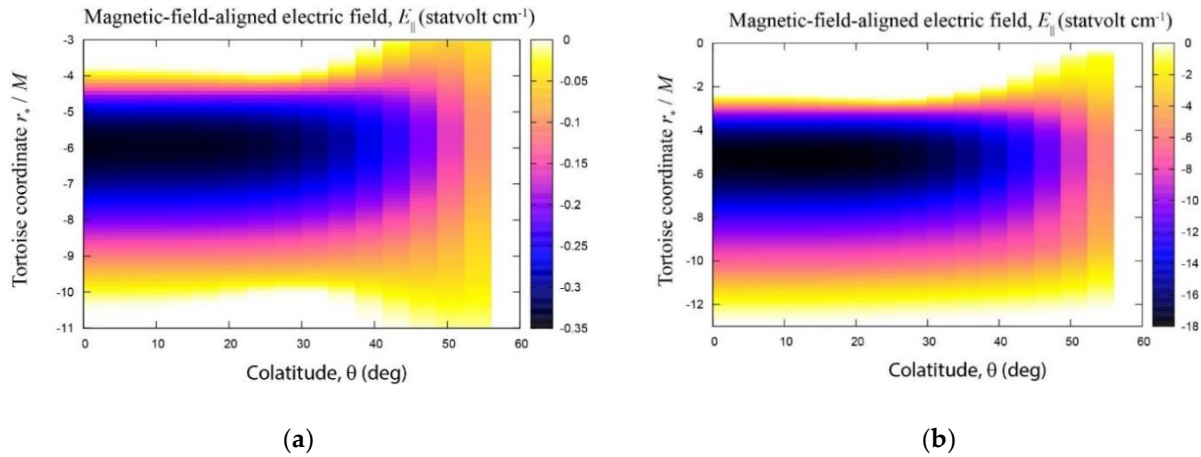

(**a**)　　　　　　　　　　　　　　　　　　　　　(**b**)

**Figure 15.** Acceleration electric field (statvolt cm$^{-1}$) on the poloidal plane. The abscissa denotes the magnetic colatitudes, $\theta$, in degrees, where 0 (i.e., the ordinate) corresponds to the magnetic axis. The ordinate denotes the dimensionless tortoise coordinate, $\eta_* = r_* / r_g$, where -∞ corresponds to the event horizon. BH spin is *a* = 0.9 *M*, and the accretion rate is set at $\dot{m}$ =1.77 × 10$^{-5}$. (**a**) The case of *M*=10$^9$ M$_\odot$. (**b**) The case of *M* = 10$^6$ M$_\odot$. From [68].

### 7.3. Results: Gap Width

Let us briefly examine how the gap width, *w*, depends on the ADAF soft-photon field as the accretion rate changes. In Figure 16, we plot the gap inner and outer boundary positions as a function of $\dot{m}$, where the ordinate is converted into the dimensionless Boyer–Lindquist radial coordinate, *r*/*M*. The left panel shows the result in the case of a billion solar-mass BH, while the right panel does that in the case of a million solar-mass BH. It follows that the gap inner boundary (solid curve, $r = r_1$) infinitesimally approaches the horizon (dot–dashed horizontal line, $r = r_H$), while the outer boundary (dashed curve, r = r$_2$) moves outward, with decreasing $\dot{m}$. At greater accretion rate, $\dot{m} > 1.8 \times 10^{-4}$ (for M = 10$^9$ M$_\odot$ case), and $\dot{m} > 10^{-3}$ (for *M* = 10$^6$ M$_\odot$ case), we fail to find stationary solutions. At smaller accretion rate, ($\dot{m} < 5.6 \times 10^{-6}$ for *M* = 10$^9$ M$_\odot$ case, and $\dot{m} < 2.3 \times 10^{-5}$ for *M* = 10$^6$ M$_\odot$ case), there is no stationary gap solution, because the pair creation becomes too inefficient to create the externally imposed current density, $j_{tot}$, even with w»M. Note that $j_{tot}$ (along each magnetic flux tube) should be constrained by a global requirement (including the dissipative region at large distances) and cannot be solved if we consider only the gap region.

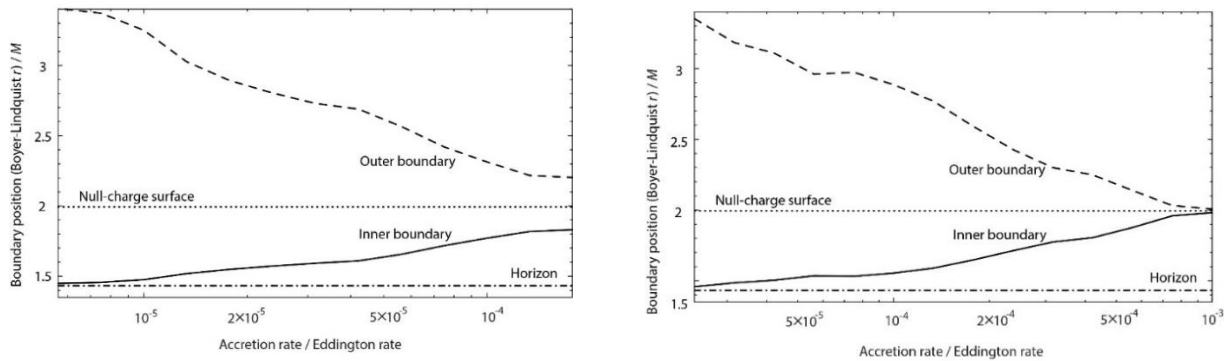

(**a**)　　　　　　　　　　　　　　　　　　(**b**)

**Figure 16.** Spatial extent of the gap along the polar axis, $\theta = 0°$, as a function of the dimensionless accretion; $a = 0.9\,M$ is assumed. Thick solid and dashed curves denote the position of the inner and the outer boundaries of the gap, respectively, in units of the gravitational radius, $r_g = M$. These boundary positions (in the ordinate) are transformed from the tortoise coordinate into the Boyer–Lindquist coordinate. The horizontal dot–dashed line shows the horizon radius, whereas the horizontal dotted line shows the null surface position. (**a**) The case of $M = 10^9 M_\odot$. (**b**) The case of $M = 10^6 M_\odot$. From [68].

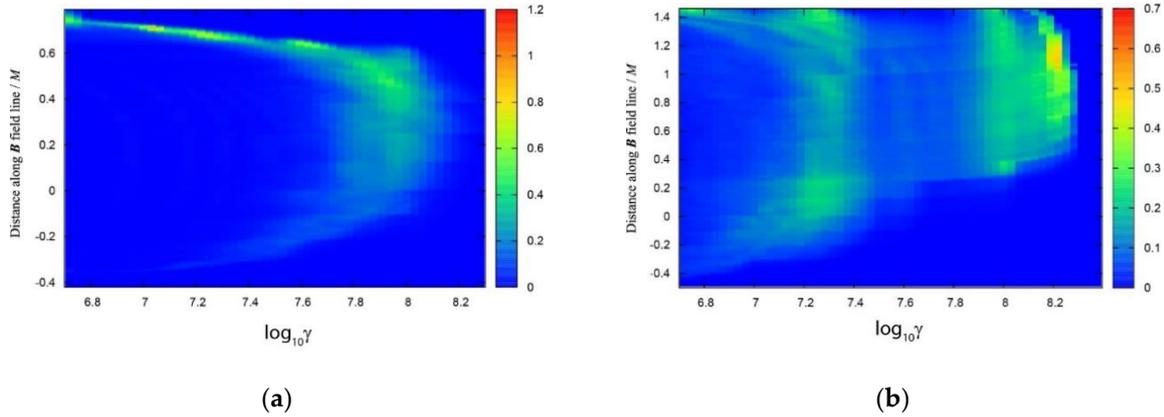

(**a**)　　　　　　　　　　　　　　　　　　(**b**)

**Figure 17.** Electron distribution function, $\gamma n_-(r,\gamma)$, of the electrons along the magnetic field line on the polar axis, $\theta = 0°$, for $a = 0.9M$ and $\dot{m} = 1.77 \times 10^{-5}$. The abscissa denotes the electron Lorentz factor, while the ordinate denotes the distance from the null surface, $s \approx r - r_0$, along the poloidal magnetic field line, where the ordinate is converted into the Boyer–Lindquist radial coordinate. (**a**) The case of $M = 10^9 M_\odot$. (**b**) The case of $M = 10^6 M_\odot$. From [63].

## 7.5. Results: Gamma-Ray Spectra

The predicted photon spectra are depicted in Figure 18 for six $\dot{m}$ values. The thin curves on the left denote the input ADAF spectra, while the thick lines on the right denote the output spectra from the gap. The left panel shows the result in the case of a billion solar-mass BH assuming the distance of $d = 10$ Mpc, while the right panel does that in the case of a million solar-mass BH with $d = 1$ Mpc. In both cases, the emitted γ-ray flux increases with decreasing $\dot{m}$, because the decreased soft photon field increases the photon–photon collision mean free path, and hence the gap width.

In the case of $M = 10^9 M_\odot$ (left panel), the spectra peak between 1 and 30 TeV, because the IC process dominates the curvature one. It is clear that the gap HE flux lies well below the detection limit of the Fermi/LAT (three thin solid curves labeled with "LAT 10 year" [139]). Nevertheless, its VHE flux appears above the CTA detection limits (dashed and dotted curves labeled with "CTA 50 h" [140]). Is is possible to argue that such a large VHE flux will be detected within one night with CTA from a nearby low-luminosity like M87, when the accretion rate coincidentally resides in the range, $6 \times 10^{-6} < \dot{m} < 3 \times 10^{-5}$.

In the case of $M = 10^6 M_\odot$ (right panel), the spectra peak between 30 GeV and 10 TeV. The IC process dominates the curvature one also for million solar-mass BHs. When the accretion rate is in the narrow range $2\times 10^{-5} < \dot{m} < 3\times 10^{-5}$, the gap emission may be detectable with CTA, particularly if the source is located in the southern sky.

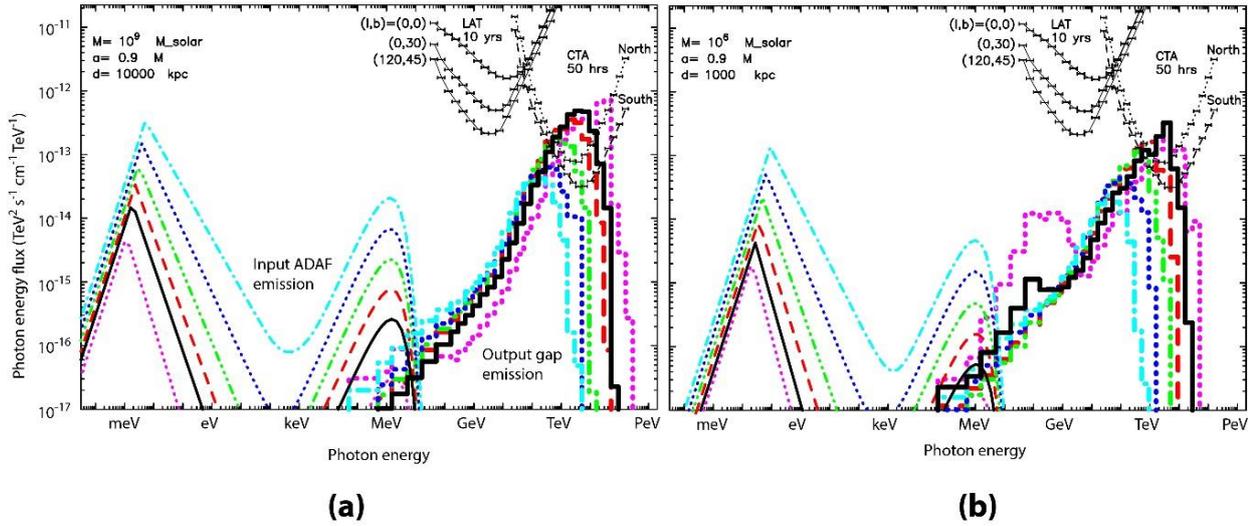

**Figure 18.** SED of the gap emission for a supermassive BH with *a* = 0.9 *M*. The thin curves denote the input ADAF spectra, while the thick lines denote the output gap spectra. The thin solid curves (with horizontal bars) denote the Fermi/LAT detection limits after 10 year of observation, while the thin dashed and dotted curves (with horizontal bars) denote the CTA detection limits after 50 h of observation. (**a**) The case of $M = 10^9 M_\odot$. The distance is set to be 10 Mpc. The cyan dot-dashed, blue dotted, green triple-dot-dashed, red dashed, black solid, and purple dotted curves correspond to $\dot{m}$ = 1.77 × 10$^{-4}$, 1.00 × 10$^{-4}$, 5.62 × 10$^{-5}$, 3.16 × 10$^{-5}$, 1.77 × 10$^{-5}$, and 5.62 × 10$^{-6}$, respectively. Magnetic field strength is assumed to be the equipartition value with the plasma accretion. (**b**) The case of $M = 10^6 M_\odot$. The distance is set to be 1 Mpc. The cyan dot-dashed, blue dotted, green triple-dot-dashed, red dashed, black solid, and purple dotted curves correspond to $\dot{m}$ = 3.16 × 10$^{-4}$, 1.00 × 10$^{-4}$, 5.62 × 10$^{-5}$, 4.21 × 10$^{-5}$, 3.16 × 10$^{-5}$, and 1.77 × 10$^{-5}$, respectively. From [63].

## 8. Discussion

### 8.1. The Case of Very Small Accretion Rate

Let us discuss the case of very small accretion rates, $\dot{m} < \dot{m}_{\text{low}}(M, a)$. The ADAF photon field peaks around eV for stellar-mass BHs and around meV for supermassive BHs. These photons are emitted from the innermost region, $r \sim R_{\min} \sim 6M$, which is located well inside the gap outer boundary when $\dot{m} \approx \dot{m}_{\text{low}}$; thus, pair production is sustained only marginally in an extended gap. However, at $\dot{m} < \dot{m}_{\text{low}}$, stationary pair production can be no longer sustained and a vacuum region develops in the entire polar funnel. In this vacuum region, migratory leptons are accelerated by the vacuum $E_\parallel$ and cascade into copious primary electrons and positrons that are accelerated in the opposite directions. The resultant emission will become inevitably non-stationary.

### 8.2. Gap Emission Versus Jet Emission

Next, let us discuss how to discriminate the gap and jet emissions. As we have seen in Sections 6 and 7, the gap gamma-ray luminosity increases with decreasing $\dot{m}$. Thus, we can predict an anti-correlation between the ADAF-emitted infrared (IR)/optical and the gap-emitted HE/VHE fluxes for stellar-mass BHs, and between the ADAF-emitted radio and the gap-emitted VHE fluxes for supermassive BHs. This forms a contrast to the standard shock-in-jet scenario, whose natural prediction will be a correlation between the radio/IR/optical and the HE/VHE fluxes (for reviews or catalogues of gamma-ray observations of AGNs, see e.g., [126,141-145]). This anti-correlation or correlation can be detectable both for stellar-mass and supermassive BHs. For stellar-mass BHs, BH transients may exhibit a correlation between IR/optical and HE/VHE fluxes during quiescence or low-hard state. For supermassive BHs, nearby low-luminosity AGNs may show an anti-correlation between submillimeter wavelength and VHE. In X-rays, the gap emission is very weak. Thus, if X-ray photons are detected, they are probably emitted from the jet or from the accretion flow.

*8.3. Comparison with Pulsar Gap Models*

Let us compare the present BH gap model with the pulsar outer (OG) gap model. In both gap models, the gap appears around the null-charge surface where the GJ charge density vanishes, as the stationary solution of the Maxwell–Boltzmann equations. Moreover, the gap width, and hence the γ-ray efficiency, which is defined by the ratio between the γ-ray and spin-down luminosities, increases with decreasing soft-photon flux. This is because the pair-production mean-free path increases with decreasing photon flux. For instance, the OG γ-ray efficiency increases with decreasing NS surface emission, or equivalently with increasing NS age [45,146]. In the same manner, the BH gap γ-ray efficiency increases with decreasing accretion rate of ADAF. In addition, electron–positron pairs are supplied by photon–photon pair production in both pulsar and BH magnetosphere, although ions can be drawn into the pulsar OG from the neutron-star surface as a space-charge limit flow [147], in the same manner as in PC models (2).

However, there are differences, as described below.

First, in a pulsar magnetosphere, the null surface appears because of the convex geometry of a dipole magnetic field. On the other hand, in a BH magnetosphere, a null surface appears due to the frame-dragging effect around a rotating BH. Thus, for the same magnetic field polarity, the sign of $E_\parallel$ is opposite.

Second, in the OG model, a soft-photon field is provided by the cooling NS thermal emission and/or the heated polar-cap thermal emission. The NS surface emissions peak in the X-ray energies. Thus, the curvature 1–10 GeV photons efficiently materialize as pairs within the gap, thereby contributing to the gap closure. However, the thermal IR photon field is much weaker than the thermal X-ray field; thus, the IC photons do not contribute to the gap closure in pulsars. On the other hand, around a BH it is the ADAF that provides the soft-photon field. For a supermassive BH, the ADAF spectrum peaks in submillimeter wavelengths. Thus, the photons having energies around 100 TeV or above efficiently materialize as pairs to close the gap. The 0.1–10 GeV curvature photons, on the other hand, do not materialize efficiently because the IR photon number flux is five to six orders of magnitude weaker than the submillimeter one.

Let us briefly compare the BH gap model with the pulsar polar-cap (PC) model. In a pulsar magnetosphere, electrons may be drawn outward as a space-charge-limited flow at the NS surface in the PC region. Thus, in a stationary gap [31,32] or in a non-stationary gap [34], γ-ray emission could be realized without pair creation within the polar cap, although pairs are indeed created via magnetic pair creation (e.g., at least at the outer boundary where $E_\parallel$ should be screened). However, in BH magnetospheres, causality prevents any plasma emission across the horizon. Thus, a gap could not be sustained without pair creation.

If we compare the stationary BH gap model with pulsar OG and PC models, it is possible to argue for the stability of gaps in a qualitative manner (§6.3 of [68]). In this review, finally we discuss the fact that the non-stationary nature of pulsar PCs [34] cannot readily be extended to the stability arguments of stationary BH gap solutions. In a pulsar PC, there exists no null surface. Thus, if a three-dimensional PC region is charge-starved in the sense $|\rho - \rho_{GJ}| \ll |\rho_{GJ}|$, a positive $-\rho_{GJ}$ leads to a negative $E_\parallel$. Consider a small-amplitude pair production taking place near the upper boundary of a PC accelerator. Created positrons migrate inwards while electrons outwards, resulting in an increased (or decreased) $\rho - \rho_{GJ}$ in the lower (or the upper) part of the pair-production site. Accordingly, $|E_\parallel|$ increases (or decreases) in the lower part (or the upper part). The increased $|E_\parallel|$ in the lower part further enhances pair production in the upper regions, because the electrons drawn from the neutron star surface will gain more energy. As a result, $|E_\parallel|$, and hence the pair production increases with time. Because of this positive feedback effect, instability sets in, as demonstrated by PIC

simulations [34]. In short, pulsar PC accelerators are inherently unstable for pair production because of $E_{\parallel} < 0$, which stems from the fact that there exists no null-charge surface in the PC region due to the relatively weak frame-dragging effect (i.e., the Lense–Thirring effect [148]) around a rotating neutron star.

On these grounds, we cannot readily conclude that the stationary BH gap solutions are unstable because of the highly time-dependent nature of the pulsar PC accelerators. Careful examinations with PIC simulations are needed for BH gaps, as performed recently by [81,83]. Since the saturated solution in recent PIC simulation is much less violently time-dependent compared to pulsar PC cases [34,81], it may be reasonable to start with stationary BH gap models as the first step.

**Funding:** This research received no external funding.

**Acknowledgments:** The author is indebted to Dr. H.-Y. Pu for providing the computed spectral data of advection-dominated accretion flows, which are used in sections 6 and 7.